\documentclass[12pt]{article}
\usepackage{color}
\usepackage{latexsym}
\usepackage{epsfig,amssymb,euscript, mathrsfs}
\usepackage{amsmath}
\newsavebox{\ns}
\newsavebox{\dbrane}
\newsavebox{\dbshort}

\def\be{\begin{equation}}
\def\ee{\end{equation}}
\def\bea{\begin{eqnarray}}
\def\eea{\end{eqnarray}}

\newcommand{\nn}{\nonumber}

\newcommand\R{\mathbb{R}}
\newcommand\Z{\mathbb{Z}}

\newcommand\C{\mathbb{C}}

\newcommand\diff{\mathrm{d}}
\newcommand{\de}{\partial}

\newcommand{\dd}{\mathrm{d}}
\newcommand{\me}{\mathrm{e}}
\newcommand{\ii}{\mathrm{i}}

\newcommand{\ex}{\mathrm{e}}
\newcommand{\vol}{\mathrm{vol}}

\newcommand{\Imag}{\mathrm{Im}\, }
\newcommand{\Real}{\mathrm{Re}\, }
\newcommand\g{\gamma}

\newcommand{\proj}{{\cal P}_-}

\newcommand{\EE}{E}

\newcommand{\GG}{\mathcal{G}}
\newcommand{\calB}{\mathcal{B}}
\newcommand{\calBhat}{\hat{\mathcal{B}}}
\newcommand{\Jhat}{\hat{J}}
\newcommand{\hatd}{\hat{\mathrm{d}}}
\newcommand{\Omegahat}{\hat{\Omega}}
\newcommand{\hh}{u}
\newcommand{\ttt}{t}
\newcommand{\bbJ}{\mathbb{J}}
\newcommand{\bbB}{\mathbb{B}}
\newcommand{\nes}{Z}
\newcommand{\wes}{W}

\numberwithin{equation}{section}       % equation numbers in each section

\begin{document}

\begin{titlepage}

\flushright{IPhT-T12/053}

\begin{center}

\today

\vskip 1.5 cm 

{\Large \bf $\mathcal{N}=2$ supersymmetric AdS$_4$ solutions of M-theory}

\vskip 1.5 cm

{Maxime Gabella$^1$, Dario Martelli$^2$, Achilleas Passias$^2$ and James Sparks$^3$\\}

\vskip 1cm

$^1$\textit{Institut de Physique Th\'eorique, CEA/Saclay
\\
91191 Gif-sur-Yvette Cedex, France\\}

\vskip 0.8cm

$^2$\textit{Department of Mathematics, King's College, London, \\
The Strand, London WC2R 2LS,  United Kingdom\\}

\vskip 0.8cm

$^3$\textit{Mathematical Institute, University of Oxford,\\
24-29 St Giles', Oxford OX1 3LB, United Kingdom\\}

\vskip 0.8cm

\textit{maxime.gabella@cea.fr, dario.martelli@kcl.ac.uk, achilleas.passias@kcl.ac.uk, sparks@maths.ox.ac.uk}

\end{center}

\vskip 1.0 cm

\begin{abstract}

\noindent We analyse the most general $\mathcal{N}=2$ supersymmetric solutions of $D=11$ supergravity 
consisting of a warped product of four-dimensional anti-de-Sitter space with a seven-dimensional Riemannian manifold $Y_7$.
 We show that the necessary and sufficient conditions for supersymmetry 
can be phrased in terms of a local $SU(2)$-structure on $Y_7$. Solutions with
non-zero M2-brane charge also admit a canonical contact structure, in terms of which 
many physical quantities can be expressed, including 
 the free energy and the scaling dimensions of operators dual to supersymmetric wrapped M5-branes. 
We show that a special class of solutions is singled out 
by imposing an additional symmetry, for which the problem reduces to solving a 
second order non-linear ODE. As well as recovering a known class of solutions, that includes
the IR fixed point of a mass deformation of the ABJM theory, 
we also find new solutions which are dual to cubic deformations. 
In particular, we find a new supersymmetric warped AdS$_4\times S^7$ solution with non-trivial four-form flux. 

\end{abstract}

\end{titlepage}

\pagestyle{plain}
\setcounter{page}{1}
\newcounter{bean}
\baselineskip18pt
\tableofcontents

\section{Introduction}

Over the last ten years 
there has been tremendous progress in our understanding of the AdS/CFT 
correspondence \cite{Maldacena:1997re} in the presence of unbroken supersymmetry. We have witnessed the discovery of
many highly non-trivial supersymmetric solutions of supergravity, together with a 
rather detailed understanding of their gauge theory duals. 
Supersymmetric solutions  with an anti-de Sitter (AdS) factor are particularly important, as they are dual
to \emph{superconformal} field theories, in an suitable limit. Comprehensive studies 
of general supersymmetric AdS geometries, in different dimensions,
 have been carried out in \cite{Martelli:2003ki,Gauntlett:2004zh,Lin:2004nb,Lust:2004ig,Gauntlett:2005ww}\footnote{The generality of the ansatz used 
in \cite{Lin:2004nb} was proven in \cite{OColgain:2010ev}.} 
and led to a number of interesting developments.
These results have all been obtained using the technique
of analysing a canonical $G$-structure 
 in order to obtain necessary and sufficient conditions for supersymmetry \cite{Gauntlett:2002sc}.
In this paper we will systematically study the most general class of $\mathcal{N}=2$ AdS$_4$ solutions of $D=11$ supergravity. 
Supersymmetric AdS$_4$ solutions of $D=11$ supergravity have been discussed  before  in the literature \cite{Lukas:2004ip, Behrndt:2005im}. 
However,  these references contain errors,  and reach incorrect conclusions that miss important classes of solutions. 

Our main motivation for studying AdS$_4$ solutions of $D=11$ supergravity in particular is that,  starting with the seminal work of \cite{Bagger:2006sk, Gustavsson:2007vu, Aharony:2008ug}, 
over the past few years there has been considerable progress in understanding the AdS$_4$/CFT$_3$ correspondence 
in M-theory. In particular, with ${\cal N}\geq 2$ supersymmetry there is good control on both sides of the duality, 
and this has led to many new examples of AdS$_4$/CFT$_3$ dualities, including infinite families, along with 
precise quantitative checks.  On the gravity side, the simplest setup
is that of Freund-Rubin AdS$_4\times\mathrm{SE}_7$  backgrounds of M-theory, where $\mathrm{SE}_7$ is a Sasaki-Einstein manifold\footnote{Particular cases with ${\cal N}>2$ include
three-Sasakian manifolds and orbifolds of the round seven-sphere.}.
%, and deformations thereof. 
These are conjectured to be dual to the theory on a 
large number $N$ of M2-branes
placed at a Calabi-Yau four-fold singularity. Rather generally,  these field theories are believed to be strongly coupled Chern-Simons-matter theories 
at a conformal fixed point. With $\mathcal{N}\geq 2$ supersymmetry the partition function of such a theory on the three-sphere 
localizes \cite{Kapustin:2009kz, Jafferis:2010un, Hama:2010av}, reducing the infinite-dimensional functional integral exactly to a finite-dimensional matrix integral. 
This can then often be computed exactly in the large $N$ limit, where $N$ is typically related to the rank of the 
gauge group, and compared to a gravitational dual computation which is purely geometric. 
Such computations have now been performed in a variety of examples \cite{Martelli:2011qj, Cheon:2011vi, Jafferis:2011zi, Amariti:2012tj}, 
with remarkable agreement on each side. 

Thus far, almost all attention has been focused on AdS$_4\times \mathrm{SE}_7$ solutions.
This is for the simple reason that very few AdS$_4$ solutions outside this class are known. 
An exception is the Corrado-Pilch-Warner solution \cite{Corrado:2001nv}, which 
describes  the infrared fixed point of a massive deformation of the maximally supersymmetric 
ABJM theory on $N$ M2-branes in flat spacetime. This solution is topologically AdS$_4\times S^7$, but 
the metric on $S^7$ is not round, and there is a non-trivial warp factor and internal four-form flux 
on the $S^7$.  This has more recently been studied in 
\cite{Klebanov:2008vq, Ahn:2009bq,  Ahn:2009sk, Jafferis:2011zi}, and in particular in the last reference the 
free energy $\mathcal{F}$ of the $\mathcal{N}=2$ superconformal fixed point was shown to match 
the free energy computed using the gravity dual solution. The Corrado-Pilch-Warner 
solution also has a simple generalization to massive deformations of $N$ M2-branes at a CY$_3\times \C$ 
four-fold singularity, where CY$_3$ denotes an arbitrary Calabi-Yau three-fold cone singularity. 

In this paper we systematically study the most general class of $\mathcal{N}=2$ 
AdS$_4$ solutions of M-theory. These have an eleven-dimensional metric which is a 
warped product of AdS$_4$ with a compact Riemannian seven-manifold $Y_7$. In order that the $SO(3, 2)$ isometry group
of AdS$_4$ is a symmetry group of the full solution, the four-form field strength necessarily has
an ``electric'' component proportional to the volume form of AdS$_4$, and 
a ``magnetic'' component which is a pull-back from $Y_7$. 
We show, with the exception of the Sasaki-Einstein case, that the geometry on $Y_7$ admits a canonical local $SU(2)$-structure, 
and determine the necessary and sufficient conditions for a supersymmetric solution 
in terms of this structure. In particular, $Y_7$ is equipped with a canonical Killing vector field $\xi$, 
which is the geometric counterpart to the $\mathtt{u}(1)$  R-symmetry of the dual ${\cal N}=2$ superconformal field theory. 

Purely magnetic solutions correspond physically to wrapped M5-brane solutions, 
and we correspondingly recover the supersymmetry equations in \cite{Gauntlett:2006ux} 
from our analysis. There is a single known solution in the literature, where $Y_7$ is an $S^4$ 
bundle over a three-manifold $\Sigma_3$ equipped with an Einstein metric of negative Ricci curvature. 
On the other hand, solutions with non-vanishing electric flux have a non-zero quantized 
M2-brane charge $N\in\mathbb{N}$, and include the Sasaki-Einstein manifold solutions as a 
special case where the magnetic flux vanishes. For the general class of solutions with non-vanishing 
M2-brane charge, we show that supersymmetry endows $Y_7$ with a canonical contact structure, 
for which the R-symmetry vector field $\xi$ is the unique Reeb vector field. A number of physical 
quantities can then be expressed purely in terms of contact volumes, including the 
gravitational free energy referred to above, and the scaling dimension of BPS operators  $\mathcal{O}_{\Sigma_5}$ dual to 
probe M5-branes wrapped on supersymmetric five-submanifolds $\Sigma_5 \subset Y_7$.
These formulae may be evaluated using topological and localization methods, allowing one 
to compute the free energy and scaling dimensions of certain BPS operators without knowing the 
detailed form of the supergravity solution. 

In our analysis we recover the Corrado-Pilch-Warner solution as a solution to our system of 
$SU(2)$-structure equations. We also show that this solution is in a subclass 
of solutions which possess an additional Killing vector field. For this subclass the 
supersymmetry conditions are equivalent to specifying a (local) K\"ahler-Einstein 
four-metric, together with a solution to a particular second order non-linear ODE. 
We show that this ODE admits a solution with the correct boundary conditions to give 
a gravity dual to the infrared fixed point of cubic deformations of $N$ M2-branes at a CY$_3\times \C$ 
four-fold singularity. In particular, when CY$_3=\C^3$ equipped with its flat metric, this 
leads to a new, smooth $\mathcal{N}=2$ supersymmetric AdS$_4\times S^7$ solution of M-theory. 

The plan of the rest of this paper is as follows. In section \ref{sec:SUSY} we analyse the general 
conditions for $\mathcal{N}= 2$ supersymmetry for a warped
AdS$_4 \times Y_7$ background of eleven-dimensional supergravity, 
reducing the equations to a local $SU(2)$-structure when $Y_7$ is not Sasaki-Einstein. 
In section \ref{sec:contact} we further elaborate on the geometry and physics of solutions
with non-vanishing electric flux, in particular showing that solutions admit a canonical contact 
structure, in terms of which various physical quantities such as the free energy may be expressed. 
This section is an expansion of material first presented in \cite{Gabella:2011sg}. Finally, in 
section \ref{sec:tauKilling} we analyse the supersymmetry conditions under the additional 
geometric assumption that a certain vector bilinear is Killing. In addition to 
recovering the Corrado-Pilch-Warner solution, we also numerically find a new class of cubic deformations of 
general CY$_3\times \C$ backgrounds. Section \ref{concsec} briefly concludes. 
A number of technical details, as well as the analysis of various special cases, are relegated to four appendices.

\vskip 5mm

{\bf \noindent Note}: shortly after submitting this paper to the arXiv, the paper  
\cite{Halmagyi:2012ic} appeared,  which contains a supersymmetric solution 
 that appears to coincide with the 
solution we present in section \ref{sec:tauKilling}.

\section{The conditions for supersymmetry}\label{sec:SUSY}

In this section we analyse the general conditions for $\mathcal{N}=2$ supersymmetry for a 
warped AdS$_4\times Y_7$ background of eleven-dimensional supergravity.

\subsection{Ansatz and spinor equations}

The bosonic fields of eleven-dimensional supergravity consist of  a metric $g_{11}$ and a three-form potential $C$ with 
four-form field strength $G=\diff C$. The signature of the metric is $(-,+,+,\ldots,+)$ and the action is
\bea\label{action}
S &=& \frac{1}{(2\pi)^8 \ell_p^9}\int  R *_{11} \mathbf{1} - \frac{1}{2}G\wedge *_{11}G - \frac{1}{6}C\wedge G \wedge G~,
\eea
with $\ell_p$ the eleven-dimensional Planck length. The resulting equations of motion are
\bea\label{EOM}
R_{MN}-\frac{1}{12}\left[G_{MPQR}G_{N}^{\  \ PQR}-\frac{1}{12}(g_{11})_{MN} G^2\right]&=&0~,\nonumber\\
\diff *_{11}G+\frac{1}{2}G\wedge G&=&0~,
\eea
where $M,N =0,\ldots,10$ denote spacetime indices.

We consider AdS$_4$ solutions of M-theory of the warped product form
\bea\label{ansatz}
g_{11} &=& \ex^{2\Delta}\left(g_{\mathrm{AdS}_4}+g_{7}\right)~,\nn\\
G&=& m\vol_4 + F~.
\eea
Here $\vol_4$ denotes the Riemannian volume form on AdS$_4$, and without loss of 
generality we 
take $\mathrm{Ric}_{\mathrm{AdS}_4}=-12 g_{\mathrm{AdS}_4}$.\footnote{\label{footing}The factor here is chosen to coincide with standard conventions in the case that $Y_7$ is a Sasaki-Einstein seven-manifold. For example, 
the AdS$_4$ metric in global coordinates then reads $g_{\mathrm{AdS}_4} =\tfrac{1}{4}( -\cosh^2\varrho\,  \dd \ttt^2 + \dd \varrho^2 + \sinh^2 \varrho\, \dd \Omega^2_2)$, where $\dd \Omega^2_2$ denotes the unit round metric on $S^2$.} In order to preserve the $SO(3,2)$ invariance 
of AdS$_4$ we take the warp factor $\Delta$ to be a function on the compact seven-manifold $Y_7$, 
and $F$ to be the  pull-back of a four-form on $Y_7$. 
The Bianchi identity $\diff G=0$ then requires that $m$ is
constant. The case in which $m\neq0$ will turn out to be quite distinct from that with $m=0$.

In an orthonormal frame, the Clifford algebra $\mathrm{Cliff}(10,1)$ is generated by gamma matrices $\Gamma_A$ satisfying 
$\{\Gamma_A,\Gamma_B\}=2\eta_{AB}$, where the frame indices
$A,B=0,\ldots, 10$, and 
$\eta=\mathrm{diag}(-1,1,\ldots,1)$, and we choose a representation with $\Gamma_0\cdots \Gamma_{10}=1$. The Killing spinor equation is
\bea\label{Killingspinoreqn}
\nabla_M\epsilon + \frac{1}{288}\left(\Gamma_M^{\ \ NPQR}-8\delta_M^N\Gamma^{PQR}\right)G_{NPQR}\, \epsilon &=&0~,
\eea
where $\epsilon$ is a Majorana spinor. We may decompose $\mathrm{Cliff}(10,1)\cong \mathrm{Cliff}(3,1)\otimes 
\mathrm{Cliff}(7,0)$ via
\bea
\Gamma_\alpha &=& \rho_\alpha\otimes 1~, \qquad \Gamma_{a+3} \ = \ \rho_5\otimes \gamma_a~,
\eea
where $\alpha, \beta= 0,1,2,3$ and $a,b=1,\ldots,7$ are orthonormal frame indices for AdS$_4$ and $Y_7$ respectively, 
$\{\rho_\alpha,\rho_\beta\}=2\eta_{\alpha\beta}$, $\{\gamma_a,\gamma_b\}=2\delta_{ab}$,
 and we have defined $\rho_5=\ii \rho_0\rho_1\rho_2\rho_3$. 
Notice that our eleven-dimensional conventions imply that $\gamma_1\cdots\gamma_7=\ii 1$.

The spinor ansatz preserving $\mathcal{N}=1$ supersymmetry in AdS$_4$ is 
\bea\label{spinoransatz}
\epsilon &=& \psi^+\otimes \ex^{\Delta/2}\chi + (\psi^+)^c\otimes \ex^{\Delta/2}\chi^c~,
\eea
where $\psi^+$ is a positive chirality Killing spinor on AdS$_4$, so $\rho_5\psi^+=\psi^+$, satisfying
\bea\label{AdS4spinor}
\nabla_\mu \psi^+ &=& \rho_\mu (\psi^+)^c~.
\eea
The superscript $c$ in (\ref{spinoransatz}) 
denotes charge conjugation in the relevant dimension, and 
the factor of $\ex^{\Delta/2}$  is included for later convenience. Substituting (\ref{spinoransatz})
into the Killing spinor equation (\ref{Killingspinoreqn}) leads to the following algebraic and differential equations for the spinor field $\chi$ on $Y_7$
\bea\label{spinoreqns}
\frac{1}{2}\gamma^n\partial_n\Delta \chi -\frac{\ii m}{6}\ex^{-3\Delta}\chi+\frac{1}{288}\ex^{-3\Delta}
F_{npqr}\gamma^{npqr}\chi+\chi^c&=&0~,\nn\\
\nabla_m\chi +\frac{\ii m}{4}\ex^{-3\Delta}\gamma_m\chi -\frac{1}{24}\ex^{-3\Delta} F_{mpqr}\gamma^{pqr}\chi 
-\gamma_m\chi^c&=&0~.
\eea
For a supergravity solution one must also solve the equations of motion (\ref{EOM}) resulting from (\ref{action}), 
as well as the Bianchi identity $\diff G=0$.

Motivated by the discussion in the introduction, in this paper we will focus on $\mathcal{N}=2$ supersymmetric AdS$_4$ solutions for which there are 
two independent solutions $\chi_1$, $\chi_2$ to (\ref{spinoreqns}). The general ${\cal N}=2$ Killing 
spinor ansatz may be written as
\bea\label{n2spinoransatz}
\epsilon & = & \sum_{i=1,2} \psi_i^+\otimes \ex^{\Delta/2}\chi_i + (\psi^+_i)^c\otimes \ex^{\Delta/2}\chi^c_i~.
\eea
In general the two Killing spinors $\psi_i^+$ on AdS$_4$ satisfy an equation of the form 
\bea
\nabla_\mu \psi_i^+ &=& \sum_{j=1}^2 W_{ij}\rho_\mu(\psi_j^+)^c~.
\label{adspinors}
\eea
Multiplying by $\bar{\psi}^+_k\rho^\mu$  on the left it is not difficult to show that $W_{ij}$ is necessarily a constant matrix.
Using the  integrability conditions of (\ref{adspinors}), 
\bea
\sum_j W_{ij} W_{jk}^* & = & \delta_{ik}~,
\eea
one can verify that, without loss of generality, by a change of basis we may take $W_{ij}=\delta_{ij}$ to be the identity matrix. 
Thus $\psi_1^+$ and $\psi_2^+$ may both be taken to satisfy (\ref{AdS4spinor}).

In this case with $\mathcal{N}=2$ supersymmetry
there is a $\mathtt{u}(1)$ R-symmetry which rotates the spinors 
as a doublet. It is then convenient to introduce 
\bea\label{chipm}
\chi_\pm &\equiv & \frac{1}{\sqrt{2}}\left(\chi_1\pm \ii \chi_2\right)~,
\eea
which will turn out to have charges $\pm2$ under the Abelian R-symmetry. 
In terms of the new basis (\ref{chipm}), the spinor equations (\ref{spinoreqns}) read
\bea\label{chispinoreqns}
\frac{1}{2}\gamma^n\partial_n\Delta \chi_\pm -\frac{\ii m}{6}\ex^{-3\Delta}\chi_\pm+\frac{1}{288}\ex^{-3\Delta}
F_{npqr}\gamma^{npqr}\chi_\pm+\chi_\mp^c&=&0~,\nn\\
\nabla_m\chi_\pm +\frac{\ii m}{4}\ex^{-3\Delta}\gamma_m\chi_\pm -\frac{1}{24}\ex^{-3\Delta} F_{mpqr}\gamma^{pqr}\chi_\pm
-\gamma_m\chi_\mp^c&=&0~.
\eea

\subsection{Preliminary analysis}\label{pre}

The condition of $\mathcal{N}=2$ supersymmetry means that the spinors $\chi_1$, $\chi_2$ in 
(\ref{n2spinoransatz}) are linearly independent. Notice that we are free to make $GL(2,\R)$ 
transformations of the pair $(\chi_1,\chi_2)$, since this leaves the spinor equations (\ref{chispinoreqns}) 
invariant. We shall make use of this freedom below.

The scalar bilinears are $\bar\chi_i\chi_j$ and $\bar\chi_i^c\chi_j$, which may equivalently be rewritten in the 
$\chi_\pm$ basis (\ref{chipm}). The differential equation in (\ref{spinoreqns}) 
immediately gives $\nabla(\bar\chi_1\chi_1)=\nabla(\bar\chi_2\chi_2)=0$, so that 
using $\R^*\times \R^*\subset GL(2,\R)$ we may without loss of generality set 
$\bar\chi_1\chi_1=\bar\chi_2\chi_2=1$. Setting $\mathcal{C}=1$ in (\ref{useful3}), the algebraic equation in (\ref{spinoreqns}) thus leads to
\bea
2\mathrm{Im}\left[ \bar{\chi_i^c}\chi_j\right]&=& - \frac{m}{3}\ex^{-3\Delta}\bar{\chi_i}\chi_j~,
\eea
where $i,j\in\{1,2\}$. We immediately conclude that
for $m\neq 0$ we have 
\bea\label{gammazero}
\mathrm{Im}\left[ \bar{\chi_1}\chi_2\right]  &=& 0~.
\eea
When $m=0$ this statement is not necessarily true. 
The case with $m=0$ and $\mathrm{Im}\left[ \bar{\chi_1}\chi_2\right]$ 
not identically zero is discussed separately in appendix  \ref{app:m0}, where we show that 
there are no regular solutions in this class. 
We may therefore take (\ref{gammazero}) to hold in all cases.

It is straightforward to analyse the remaining scalar bilinear equations. In particular, 
$\mathrm{Re}\left[ \bar\chi_1\chi_2\right]$ is constant, and 
using the remaining $GL(2,\R)$ freedom one can without loss of generality
set $\mathrm{Re}\left[ \bar\chi_1\chi_2\right]=0$.\footnote{In the 
special case that $\mathrm{Re}\left[\bar\chi_1\chi_2\right]=1$ one can show that $\chi_1=\chi_2$, which 
in turn leads to only $\mathcal{N}=1$ supersymmetry.}  In the $\chi_\pm$ basis (\ref{chipm}) we may then summarize the results 
of this analysis as
\bea\label{scalars}
\bar\chi_+\chi_+ & = & 1 \ = \  \bar\chi_-\chi_-~, \qquad \quad\  \bar\chi_+\chi_- \ = \ 0~,\nonumber\\
\bar\chi_+^c\chi_+ & \equiv & S \  = \ (\bar\chi_-^c\chi_-)^*~, \qquad \bar\chi_+^c\chi_- \ = \ -\ii \zeta~.
\eea
Here $S$ is a complex function on $Y_7$, while it is convenient to define $\zeta$ to be the real function
\bea\label{zeta}
\zeta &\equiv & \frac{m}{6}\ex^{-3\Delta}~.
\eea
Notice that in the $m=0$ limit we have $\zeta\equiv 0$, while 
for $m\neq0$ instead $\zeta$ is nowhere zero. We also define the one-form bilinears
\bea\label{vectors}
K & \equiv & \ii \bar\chi_+^c\gamma_{(1)}\chi_-~, \qquad  L \ \equiv \ \bar\chi_-\gamma_{(1)}\chi_+~,\nonumber\\
\bar\chi_+\gamma_{(1)}\chi_+ &\equiv & -P \ = \ - \bar\chi_-\gamma_{(1)}\chi_-~.
\eea
Here we have denoted $\gamma_{(n)}\equiv \frac{1}{n!}\gamma_{m_1\cdots m_n}\diff y^{m_1}\wedge \cdots \wedge \diff y^{m_n}$. 
{\it A priori} notice that $K$ and $L$ are complex, while $P$ is real.

\subsection{The R-symmetry Killing vector}

The spinor equations (\ref{chispinoreqns}) imply that 
\bea\label{K4eqn}
2 \mathrm{Im}\,  K &=& \diff \mathrm{Im}\left[\bar\chi_1\chi_2\right] \ = \ 0~,
\eea
where we have used (\ref{gammazero}). Thus in fact $K$ is real, and it is then straightforward to 
show that $K$ is a Killing one-form for the metric $g_{7}$ on $Y_7$, and hence that the dual 
vector field $\xi\equiv\  g^{-1}_{7}(K,\cdot\, )$ is a Killing vector field. More precisely, 
one computes
\bea\label{Lie}
\nabla_{(m} K_{n)} &=& -2\ii\, \mathrm{Im}\left[\bar\chi_1\chi_2\right] g_{7\, mn} \ = \ 0~.
\eea
Using the Fierz identity (\ref{Fierz}) one computes the square norm
\bea\label{norm}
\|\xi\|^2 &\equiv & g_{7}(\xi,\xi) \ = \ |S|^2 + \zeta^2~.
\eea
In particular when $m\neq 0$ we see from (\ref{zeta}) that $\xi$ is nowhere zero, and thus defines a one-dimensional foliation of $Y_7$.
In the case that $m=0$ this latter conclusion is no longer true in general, as we will show in section \ref{m0limit} via a counterexample.

The algebraic equation in (\ref{chispinoreqns}) 
leads immediately to $\mathcal{L}_\xi \Delta=0$, and using both equations in (\ref{chispinoreqns}) 
one can show that
\bea
\diff (\ex^{3\Delta}\, \bar\chi_+^c\gamma_{(2)}\chi_-) &=& - \ii \xi\lrcorner F~.
\eea
It follows that
\bea
\mathcal{L}_\xi F &=& \diff (\xi\lrcorner F)+ \xi\lrcorner\diff F \ = \ 0~,
\eea
provided the Bianchi identity $\diff F=0$ holds.\footnote{In fact this is implied by supersymmetry when $m\neq 0$, as we will show shortly 
in section \ref{sec:EOM}.}  Thus $\xi$ preserves all of the bosonic fields. 

One can also show that
\bea
\mathcal{L}_{\xi} \chi_\pm &=& \pm 2\ii\, \chi_\pm~,\label{charge2}
\eea
so that $\chi_\pm$ have charges $\pm 2$ under $\xi$. Perhaps the easiest 
way to prove this is to use the remaining non-trivial scalar bilinear equation
\bea\label{Sbilinear}
\ex^{-3\Delta}\diff(\ex^{3\Delta}S) &=& 4L~,
\eea
to show that
\bea\label{Scharge}
\mathcal{L}_\xi S &=& 4\ii S~.
\eea
Since $\xi$ preserves all of the bosonic fields, we may take the Lie derivative of the spinor equations 
(\ref{chispinoreqns}) to conclude that $\mathcal{L}_{\xi}\chi_\pm$ satisfy the \emph{same} equations, 
and hence $\mathcal{L}_\xi\chi_\pm$ are linear combinations of $\chi_\pm$. 
The Lie derivatives of the scalar bilinears, in particular (\ref{Scharge}), then fix (\ref{charge2}).\footnote{More precisely, 
this argument is valid provided $S$ is not identically zero. However, when $S=0$
we necessarily reduce to the Sasaki-Einstein case, as shown in appendix \ref{app:SE}. In that 
case (\ref{charge2}) also holds.}
We thus identify $\xi$ 
as the canonical vector field dual to the R-symmetry of the $\mathcal{N}=2$ SCFT.

\subsection{Equations of motion}\label{sec:EOM}

Given our ansatz, the equation of motion and Bianchi identity for $G$ reduce to
\bea\label{Feqns}
\diff\left(\ex^{3\Delta}\star F\right) &=& - m F~, \qquad \diff F \ = \ 0~,
\eea
where $\star $ denotes the Hodge star operator on $Y_7$. We begin by showing 
that supersymmetry implies the equation of motion, and 
that for $m\neq 0$ it also implies the Bianchi identity. 

The imaginary part of the bilinear equation for the three-form $\bar\chi_+^c\gamma_{(3)}\chi_-$ 
leads immediately to
\bea\label{mF}
m F &=&6\, \diff \left(\ex^{6\Delta}\mathrm{Im}\left[ \bar\chi_+^c \gamma_{(3)}\chi_-\right]\right)~.
\eea
Thus for $m\neq 0$ we deduce that $F$ is closed. On the other hand, the bilinear equation for the two-form $\bar\chi_+\gamma_{(2)}\chi_+$
\bea\label{*Feqn}
\ex^{3\Delta}\star F &=& \diff \left( \ii\, \ex^{6\Delta}\bar{\chi}_+\gamma_{(2)}\chi_+\right)  - 6\ex^{6\Delta}\mathrm{Im}\left[ \bar\chi_+^c \gamma_{(3)}\chi_-\right]~,
\eea
 gives, via taking the 
exterior derivative,
\bea
\diff \left(\ex^{3\Delta}\star F\right) &=& -6\diff\left(\ex^{6\Delta}\mathrm{Im}\left[\bar\chi_+^c\gamma_{(3)}\chi_-\right]\right) \ = \  -mF~,
\eea
where in the second equality we have combined with  equation (\ref{mF}). We thus see that supersymmetry implies the equation of motion in (\ref{Feqns}).

Finally,  using the integrability results of \cite{Gauntlett:2002fz} one can now show that the Einstein equation is automatically
implied as an integrability condition for the supersymmetry conditions, once the $G$-field
equation and Bianchi identity are imposed. In particular, note that the eleven-dimensional one-form bilinear 
$k\equiv \bar\epsilon\Gamma_{(1)}\epsilon$ is dual to a timelike Killing vector field, 
as discussed in \cite{Gabella:2011sg} and later in section \ref{sec:M5}. We thus conclude

\begin{quote}
\emph{For the class of $\mathcal{N}=2$ supersymmetric $\mathrm{AdS}_4$ solutions of the form (\ref{ansatz}), 
supersymmetry and the Bianchi identity $\diff F=0$ imply the equations of motion 
for $G$ and the Einstein equations. Moreover, when $m\neq0$ the Bianchi identity 
$\diff F=0$ is also implied by supersymmetry.}
\end{quote}

Note that similar results were obtained in both \cite{Gauntlett:2005ww} and \cite{Gauntlett:2004zh}.
In fact we will see in section \ref{m0limit} that the $m=0$ supersymmetry conditions also imply
the Bianchi identity, although the arguments we have presented so far do not allow us to 
conclude this yet. 

\subsection{Introducing a canonical frame}\label{sec:frame}

Provided the three real one-forms $K, \mathrm{Re}\, S^*L, \mathrm{Im}\, S^*L$ 
defined in (\ref{vectors}) are linearly independent, we may use them 
to in turn define a canonical orthonormal three-frame $\{\EE_1,\EE_2,\EE_3\}$\footnote{We use 
$S^*L$ here, as opposed to $L$, since $S^*L$ is invariant under the R-symmetry generated by $\xi$.
In particular, from the definitions in (\ref{vectors}),
and using (\ref{charge2}), (\ref{Scharge}), we have that ${\cal L}_\xi K = {\cal L}_\xi (S^* L)=0$.}.
More precisely, if these three one-forms are 
linearly independent at a point in $Y_7$,  the stabilizer group $\GG\subset \mathrm{Spin}(7)$ of 
the pair of spinors $\chi_\pm$ at that point is $\GG\cong SU(2)$, giving a natural identification 
of the tangent space with $\C^2\oplus \R \EE_1\oplus 
\R \EE_2 \oplus \R \EE_3$. Here the $SU(2)$ structure group acts 
on $\C^2$ in the vector representation. If this is true in an open set, it will turn out that 
we may go further and also introduce three canonical coordinates 
associated to the three-frame $\{\EE_1,\EE_2,\EE_3\}$.\footnote{Just from 
group theory it must be the case that the one-form $P$
in (\ref{vectors}) is a linear combination of $K$ and $S^*L$, and 
indeed one finds that $\zeta P = K+\mathrm{Im}\, SL^*$.}

We study the case that $K, \mathrm{Re}\, S^*L, \mathrm{Im}\, S^*L$  are linearly 
\emph{dependent} in appendix \ref{app:SE}. In particular, for $m\neq0$ 
we conclude that at least one of $S=0$ or 
$\|\xi\| =1$ holds at such a point. If this is the case over the whole of $Y_7$ 
(or, using analyticity and connectedness, if this is the case on any open subset of $Y_7$)  
then we show that $Y_7$ is necessary Sasaki-Einstein with $F=0$. 
Of course, in general the three one-forms can become linearly 
dependent over certain submanifolds of $Y_7$, 
and here our orthonormal frame and coordinates will break down.\footnote{This is sometimes 
referred to as a dynamical $SU(2)$ structure.} By analogy with the corresponding 
situation for AdS$_5$ solutions of type IIB string theory studied in \cite{Gabella:2009wu}, 
one expects this locus to be the same as the subspace where a pointlike M2-brane 
is BPS, and thus correspond to the Abelian moduli space of the dual CFT, although we will not pursue this comment further here.

Returning to the generic case in which $K, \mathrm{Re}\, S^*L, \mathrm{Im}\, S^*L$  are 
linearly independent in some region, 
 we may begin by introducing a coordinate $\psi$ along the orbits of the Reeb vector field $\xi$, so that
\bea
\xi &\equiv & 4\frac{\partial}{\partial \psi}~.
\eea
The equation (\ref{Scharge}) then implies that we may write
\bea\label{Scoords}
S &=& \ex^{-3\Delta}\rho\, \ex^{\ii (\psi -\tau)}~.
\eea
This defines the real functions $\rho$ and $\tau$, which will serve as two additional coordinates 
on $Y_7$. The factor of $\ex^{-3\Delta}$ has been included partly for convenience, 
and partly to agree with conventions defined in \cite{Gauntlett:2006ux} that we will recover from 
the $m=0$ limit in section \ref{m0limit}.
Using (\ref{Sbilinear}) together with the Fierz identity (\ref{Fierz}), one can then check that 
\bea
\EE_1 &\equiv &  \frac{1}{\|\xi\|} K \ = \  \frac{1}{4}\|\xi\|(\diff\psi + \mathcal{A})~,\nonumber\\
\EE_2 & \equiv &  \frac{1}{|S|\sqrt{1-\|\xi\|^2}}\, \mathrm{Re} \, S^*L \ = \ \frac{\ex^{-3\Delta}}{4\sqrt{1-\|\xi\|^2}}\diff \rho~,\nonumber\\
\EE_3 &\equiv &  \frac{|S|}{\zeta\|\xi\|\sqrt{1-\|\xi\|^2}}\left(K- \frac{\|\xi\|^2}{|S|^2}\mathrm{Im}\, S^*L\right) \ = \ \frac{|S|\|\xi\|}{4\zeta\sqrt{1-\|\xi\|^2}}(\diff\tau+\mathcal{A})~,\label{frame}
\eea
are orthonormal. Here $\mathcal{A}$ is a local one-form that is basic for the foliation 
defined by the Reeb vector field $\xi$, {\it i.e.} $\mathcal{L}_\xi \mathcal{A}=0$, 
$\xi\lrcorner\mathcal{A}=0$. Note here that 
\bea\label{Reeblength}
\|\xi\|^2 &\equiv & g_{Y_7}(\xi,\xi) \ = \ \zeta^2+ |S|^2 \ = \ \zeta^2 +\ex^{-6\Delta}\rho^2 \ =\ \frac{\ex^{-6\Delta}}{36} (m^2 + 36 \rho^2)~,
\eea
is the square length of the Reeb vector field. 
The metric on $Y_7$ may then be written as
\bea\label{metric}
g_7 &=& g_{SU(2)} + \EE_1^2 + \EE_2^2+ \EE_3^2~.\eea
We may now in turn introduce an orthonormal frame $\{e_a\}_{a=1}^4$ 
for $g_{SU(2)}$, and define the $SU(2)$-invariant two-forms
\bea
J &\equiv & J_3 \ \equiv \ e_1\wedge e_2 + e_3\wedge e_4~,\nonumber\\
\Omega & \equiv & J_1+\ii J_2 \ \equiv \ (e_1+\ii e_2)\wedge (e_3+\ii  e_4)~.
\eea
Of course, such a choice is not unique -- we are free to make $SU(2)_R$ rotations, 
under which $J_I$, $I=1,2,3$, transform as a triplet, 
where the structure group is $\GG\cong SU(2)=SU(2)_L$, and 
$\mathrm{Spin}(4)\cong SU(2)_L\times SU(2)_R$ is the spin group 
associated to $g_{SU(2)}$.

\subsection{Necessary and sufficient conditions}\label{sec:NS}

Any spinor bilinear may be written in terms of $\EE_i$, $J_I$, having chosen 
a convenient basis\footnote{Notice that using the definition of the two-form bilinears in (\ref{deftwobil}),
and the fact that ${\cal L}_\xi E_i=0$, we see that also 
 the $J_I$ are invariant under $\xi$, namely ${\cal L}_\xi J_I =0$.} for the $J_I$. 
Having solved for the one-forms in (\ref{frame}),  the remaining differential conditions arising from 
$k$-form bilinears, for all $k\leq 3$, then be shown to reduce 
(after some lengthy computations) 
to the following system of three equations
\begin{equation}\label{SUSYs}
\boxed{
\begin{array}{rcl}
\ex^{-3\Delta}\diff \left[\|\xi\|^{-1}\left(\frac{m}{6} E_1 + \ex^{3\Delta}|S|\sqrt{1-\|\xi\|^2}E_3\right)\right] &=& 2 J_3 - 2\|\xi\|E_2\wedge E_3 ~,  \\
\diff (\|\xi\|^2 \ex^{9\Delta} J_2 \wedge E_2) - \ex^{3\Delta} |S| \diff (\|\xi\| \ex^{6\Delta} |S|^{-1} J_1 \wedge E_3 ) &= &0~,  \\
\diff( \ex^{6\Delta} J_1 \wedge E_2 ) + \ex^{3\Delta} |S| \diff (\|\xi\| \ex^{3\Delta} |S|^{-1} J_2 \wedge E_3 ) &=& 0~,
\end{array}
}
\end{equation}
where in addition the flux is determined by the equation
\bea
\diff ( \ex^{6\Delta} \sqrt{1-\|\xi\|^2} J_2) &=& - \ex^{3\Delta} \star F - 6\ex^{6\Delta}\mathrm{Im}\left[ \bar\chi_+^c \gamma_{(3)}\chi_-\right]~.
\eea
Notice this is the same equation (\ref{*Feqn}) we already used in proving that the equation of motion 
for $G$ follows from supersymmetry.
The bilinear on the right hand side is given in terms of our frame by 
\bea
\mathrm{Im}\left[ \bar\chi_+^c \gamma_{(3)}\chi_-\right] &= &  
%\frac{1}{\zeta \sqrt{1-\|\xi\|^2}} \mathrm{Re} \Big[ \left( (\zeta^2 -1) K   + \mathrm{Im} S^*L 
%+ \ii  \zeta\mathrm{Re}S^*L \right) \wedge \Omega \Big] \nonumber\\[2mm] & = & 
|S| J_2 \wedge E_2 - \frac{1}{\|\xi\|}J_1 \wedge (\zeta\sqrt{1-\|\xi\|^2}E_1 + |S|E_3) ~.
\eea
One can invert the expression for the flux using these equations to obtain
\begin{equation}\label{flux}
\boxed{ F \ = \ \frac{1}{\|\xi\|}E_1\wedge \diff\left(\ex^{3\Delta}\sqrt{1-\|\xi\|^2}J_1\right)-m\frac{\sqrt{1-\|\xi\|^2}}{\|\xi\|}J_1\wedge E_2\wedge E_3~.}
\end{equation}
Notice that although we have written these equations in terms of the three real functions $|S|$, $\|\xi\|$ and $\zeta$, in 
fact they obey (\ref{Reeblength}), where $\zeta$ is given by (\ref{zeta}). Regarding $\rho$ as a coordinate, there is then really only one independent function 
in these equations, which may be taken to be the warp factor $\Delta$. We also note that the connection one-form 
$\mathcal{A}$, defined via the orthonormal frame (\ref{frame}), has curvature determined by the first equation 
in (\ref{SUSYs}), giving
\bea
\diff\mathcal{A} &=&  \frac{4m\ex^{-3\Delta}}{3\|\xi\|^2}\left[J_3 + \left(3\|\xi\|-\frac{4}{\|\xi\|}\right)E_2\wedge E_3\right]~.
\eea

\subsubsection*{Proof of sufficiency}

It is important to stress that the set of equations (\ref{SUSYs}), where the three-frame $\{E_i\}_{i=1}^3$ is given by 
(\ref{frame}), are both necessary and sufficient for a supersymmetric solution. In order 
to see this, we recall that our $SU(2)$ structure can be thought of in terms of the two $SU(3)$ 
structures defined by the spinors $\chi_+$, $\chi_-$ (or equivalently $\chi_1$, $\chi_2$). 
Each of these determines 
a real vector $\mathcal{K}_\pm\equiv \bar\chi_\pm\gamma_{(1)}\chi_\pm$, real two-form $\mathcal{J}_\pm\equiv -\ii\bar\chi_\pm\gamma_{(2)}\chi_\pm$, 
and complex three-form $\Omega_\pm\equiv\bar\chi_\pm^c\gamma_{(3)}\chi_\pm$, where recall that also
$\bar\chi_+\chi_+=\bar\chi_-\chi_-=1$. In fact $\mathcal{K}_+=-\mathcal{K}_- = -P$, so that the 
vectors determined by each $SU(3)$ structure are equal and opposite,
and $(\mathcal{J}_\pm,\Omega_\pm)$ determine two $SU(3)$ structures on 
the transverse six-space $P^\perp$. 

Let us now turn to the Killing spinor equations in (\ref{chispinoreqns}). We have two copies 
of these equations, one for each $SU(3)$ structure determined by the spinors $\chi_\pm$. 
We shall refer to the first equation in (\ref{chispinoreqns}) as the algebraic Killing spinor equation 
(it contains no derivative acting on the spinor itself). Using this notice that we may eliminate 
the $\chi_\mp^c$ term in the second equation, in order to get an equation linear in $\chi_\pm$; we shall refer
the resulting equation as the differential Killing spinor equation.
For each choice of $\pm$, the latter may be phrased in terms of a 
generalized connection $\nabla_\pm^{(T)}$, where $\nabla$ is the  Levi-Civita connection. 
The intrinsic torsion is then defined as $\tau_\pm\equiv \nabla_\pm^{(T)}-\nabla$ for each 
$SU(3)$ structure, and may be decomposed into irreducible $SU(3)$-modules 
as a section of $\Lambda^1\otimes \Lambda^2$.  Since $\Lambda^2\cong \mathtt{so}(7)=\mathtt{su}(3)\oplus 
\mathtt{su}(3)^\perp$, the intrinsic torsion may be identified as a section of $\Lambda^1\otimes \mathtt{su}(3)^\perp$. 
It is then a fact that the exterior derivatives of $\mathcal{K}_\pm$, $\mathcal{J}_\pm$, $\Omega_\pm$
 determine completely the intrinsic torsion $\tau_\pm$ -- the identifications of the irreducible 
modules are given explicitly in section 2.3 of \cite{Behrndt:2005im}. Our equations (\ref{SUSYs}) 
certainly imply the exterior derivatives of both $SU(3)$ structures, since they imply 
the exterior derivatives of all $k$-form bilinears, for $k\leq 3$.  It follows that 
from our supersymmetry equations we could (in principle) construct  both $\tau_\pm$, 
and hence write down connections $\nabla_\pm^{(T)}=\nabla+\tau_\pm$ which 
preserve each spinor, so $\nabla_\pm^{(T)}\chi_\pm=0$. 
In other words, our conditions then imply the differential Killing spinor equations 
for each of the $\mathcal{N}=2$ supersymmetries.

For the algebraic Killing spinor equation, note first that $\{\chi,\gamma_m\chi \mid m=1,\ldots,7\}$ 
forms a basis for the spinor space for each $\chi=\chi_\pm$. Thus in order for 
the algebraic equation to hold, it is sufficient that the bilinear equations resulting 
from the contraction of the algebraic Killing spinor equation with $\bar\chi$ and $\bar\chi\gamma_m$ hold, 
where $\chi$ is either of $\chi_\pm$. However, this is precisely how the identities 
in appendix \ref{app:useful} were derived. We thus find that the $\chi_+$ 
algebraic Killing spinor equation in (\ref{chispinoreqns}) is implied by the two zero-form equations
\bea
-\frac{m}{3}\ex^{-3\Delta} + 2\mathrm{Im}\, \bar{\chi}_+\chi_-^c &=& 0~,\nn\label{algebraiczero}\\
\diff\Delta\lrcorner\mathcal{K}_+ + \frac{1}{6}\ex^{-3\Delta}\bar{\chi}_+\gamma_{(4)}\chi_+ \lrcorner F &=& 0~,
\eea
and the one-form equations
\bea
\diff\Delta + \frac{1}{6}\ex^{-3\Delta}\mathcal{J}_+\lrcorner \star F & = & 0~,\nn\\\label{algebraicone}
\frac{m}{3}\ex^{-3\Delta} P - 2K + \mathcal{J}_+(\diff\Delta) - \frac{1}{6}\ex^{-3\Delta} (\ii \bar\chi_+\gamma_{(3)}\chi_+)\lrcorner F &=& 0~,
\eea
with similar equations for $\chi_-$. Notice that the first equation in (\ref{algebraiczero}) is simply the 
scalar bilinear in (\ref{scalars}) which determines $\zeta=(m/6)\ex^{-3\Delta}$. The reader can find explicit expressions for the real two-form $\mathcal{J}_+$ 
and three-form $\ii\bar{\chi}_+\gamma_{(3)}\chi_+$, in terms of the $SU(2)$-structure, in appendix \ref{app:structures}. 
Using these expressions, one can show that (\ref{SUSYs}) imply the remaining scalar equation in (\ref{algebraiczero}) and both of the equations in (\ref{algebraicone}), thus 
proving that our differential system (\ref{SUSYs}) also implies the algebraic Killing spinor equations. 
The computation is somewhat tedious, and is best done by splitting the equations (\ref{SUSYs}) into 
components under the $1+1+1+4$ decomposition 
implied by the three-frame (\ref{frame}). This decomposition is performed explicitly in section \ref{sec:reduction}.
In the second equation  in (\ref{algebraiczero}) we note that each term is in fact separately zero. 
We also note that the first equation in (\ref{algebraicone}) may be rewritten as
\bea
\mathcal{J}_+\lrcorner \diff (\ex^{6\Delta} {\cal J}_+ ) & = & \diff \left(\ex^{6\Delta}(1-\tfrac{3}{2}|S|^2)\right)~.
\label{leeform}
\eea
The left hand side is essentially the \emph{Lee form}
associated to the $SU(3)$-structure defined by $\chi_+$.\footnote{Therefore (\ref{leeform}) has 
the geometrical interpretation that the transverse six-dimensional space $P^\perp$ is conformally balanced.} 

To conclude, we have shown that (\ref{SUSYs}) are necessary and sufficient to satisfy the original Killing spinor equations (\ref{chispinoreqns}).

\subsection{M5-brane solutions: $m=0$}\label{m0limit}

It is straightforward to take the $m=0$ limit of the frame (\ref{frame}), 
differential conditions (\ref{SUSYs}), and flux $F$ given by (\ref{flux}). 
Denoting $\hat{w}=\ex^{\Delta}E_3$, $\hat{\rho}=\ex^{\Delta}E_2$, $\hat{J}_I=\ex^{2\Delta}J_I$ and $\lambda=\ex^{-2\Delta}$ we obtain the metric
\bea
\lambda^{-1}g_7 &=& \widehat{g_{SU(2)}} + \hat{w}^2 + \frac{1}{16}\lambda^2\left(\frac{\diff\rho^2}{1-\lambda^3\rho^2}+\rho^2\diff\psi^2\right)~,
\eea
with corresponding differential conditions
\bea\label{m0diff}
\diff\left(\lambda^{-1}\sqrt{1-\lambda^3\rho^2}\hat{w}\right)&=& 2\lambda^{-1/2}\hat{J}_3 + 2\rho \lambda\hat{w}\wedge \hat{\rho}~,\nonumber\\
\diff\left(\lambda^{-3/2}\hat{J}_1\wedge \hat{w}-\rho \hat{J}_2\wedge\hat{\rho}\right)&=& 0~,\nn\\
\diff\left(\hat{J}_2\wedge \hat{w}+\lambda^{-3/2}\rho^{-1}\hat{J}_1\wedge\hat{\rho}\right)&=& 0~.
\eea
The flux $F$ in (\ref{flux}) then becomes
\bea\label{Fm0}
F &=& \frac{1}{4}\diff\psi \wedge \diff\left(\lambda^{-1/2}\sqrt{1-\lambda^3\rho^2}\hat{J}_1\right)~.
\eea
These expressions precisely coincide with those in section 7.2 of \cite{Gauntlett:2006ux}. 
Of course, this is an important cross-check of our general formulae. 

Notice that the Bianchi identity for $F$ is satisfied automatically from the expression in (\ref{Fm0}).
In fact for the general $m=0$ class of geometries the Bianchi identity and equation of motion for $F$ read
\bea
\diff F &=& 0~, \qquad \diff \left(\ex^{3\Delta}\star F\right) \ = \ 0~.
\eea
Defining the conformally related metric $\tilde{g}_7 = \ex^{-6\Delta} g_7$, the equation of motion 
for $F$ becomes $\diff\, \tilde{\star} F=0$. It follows that $F$ is a harmonic four-form on 
$(Y_7,\tilde{g})$. In particular, imposing also flux quantization we see that $F$ defines a non-trivial cohomology class in 
$H^4(Y_7;\Z)$, which we may associate with the M5-brane charge of the solution. 

When $m=0$ there is no ``electric'' component of the four-form flux $G$, and 
these AdS$_4$ backgrounds have the physical interpretation of being created by 
wrapped M5-branes. Indeed, as we shall see in section \ref{sec:contact}, 
when $m\neq0$ there is always a non-zero quantized M2-brane charge $N\in\mathbb{N}$, 
with the supergravity description being valid in a large $N$ limit. 
The supergravity free energy then scales universally as $N^{3/2}$. 
One would expect the free energy of the M5-brane solutions, sourced by the 
internal ``magnetic'' flux $F$, to scale as $N^3$, where the cohomology 
class in $H^4(Y_7;\Z)$ defined by $F$ scales as $N$.
However, the lack of 
a contact structure in this case (see below) means that a proof would look rather different 
from the analysis in section \ref{sec:contact}.

In section 9.5 of \cite{Gauntlett:2006ux} the authors found a solution within the $m=0$ class, 
solving the system (\ref{m0diff}),
describing the near-horizon limit of M5-branes wrapping a Special Lagrangian three-cycle 
$\Sigma_3$. In fact this is the eleven-dimensional uplift of a seven-dimensional solution 
found originally in reference \cite{PS}.
The internal seven-manifold $Y_7$ takes the form of an $S^4$ fibration 
over $\Sigma_3$, where the latter is endowed with an Einstein metric of constant negative 
curvature. As one sees explicitly from the solution, the R-symmetry vector field $\partial_\psi$ 
acts on $S^4\subset \R^5=\R^3\oplus \R^2$ by rotating the $\R^2$ factor in the latter decomposition. 
In particular, there is a fixed copy of $S^2$, implying that $\partial_\psi$ does \emph{not} 
define a one-dimensional foliation in this $m=0$ case. Notice this also implies there 
cannot be any compatible global contact structure, again in contrast with the $m\neq0$ 
geometries. The flux $F$ generates the cohomology group $H^4(\Sigma_3\times S^4;\R)\cong \R$.

As far as we are aware, the solution in section 9.5 of \cite{Gauntlett:2006ux}
is the only known solution in this class. It would certainly be very interesting to know if there are 
more AdS$_4$ geometries sourced only by M5-branes.

\subsection{Reduction of the equations in components}
\label{sec:reduction}

In this section we further analyse the system of supersymmetry equations (\ref{SUSYs}), 
extracting information from each component under the natural $1+1+1+4$ decomposition 
implied by the three-frame (\ref{frame}). Since we have dealt with the $m=0$ equations in the previous 
section, we henceforth take $m\neq 0$ in the remainder of the paper.

We begin by defining the one-form
\bea\label{calB}
\calB & \equiv & \frac{\|\xi\|^2}{\zeta^2}\left(\diff\tau+ \mathcal{A}\right)~,
\eea
which appears in the frame element $E_3$ in (\ref{frame}), so that
\bea
E_3 &=& \frac{|S|\zeta}{4\|\xi\|\sqrt{1-\|\xi\|^2}}\calB~,
\eea
and further decompose
\bea
\calB &\equiv & \calB_\tau \diff\tau+\calBhat~,
\eea
where $\partial_\tau\lrcorner \calBhat=0$. Since also $E_1$ and $E_2$ are orthogonal 
to $\calB$, it follows that $\calBhat$ is a linear combination of $e_a$, $a=1,2,3,4$, the 
orthonormal frame for the four-metric $g_{SU(2)}$ in (\ref{metric}).
It is also convenient to rescale the latter four-metric, together with its $SU(2)$ structure, via 
\bea
\Jhat_I &\equiv & \frac{4}{\zeta} J_I~, \qquad I=1,2,3~,
\eea
so that correspondingly $\widehat{g_{SU(2)}} = (4/\zeta)g_{SU(2)}$.\footnote{This scaling is different 
from the scaling used in section \ref{m0limit}, where $m=0$.} Notice this makes 
sense only when $m\neq 0$, so that $\zeta$ is nowhere zero. 

Given the  coordinates $(\psi,\tau,\rho)$ defined via (\ref{frame}), it is then 
natural to decompose the exterior derivative as
\bea
\diff &=& \diff\psi\wedge \frac{\partial}{\partial\psi} + \diff\tau\wedge \frac{\partial}{\partial\tau} 
+ \diff\rho\wedge\frac{\partial}{\partial \rho} + \hatd~,
\eea 
where from now on hatted expressions will (essentially) denote four-dimensional quantities. 
We may then decompose the exterior derivatives and forms in the supersymmetry equations 
(\ref{SUSYs}) under this natural $1+1+1+4$ splitting. 

Beginning with the first equation in (\ref{SUSYs}), the utility of the definition (\ref{calB}) is that this first supersymmetry equation becomes 
simply
\bea\label{SUSY1}
\diff\calB &=& 2\hat{J}_3 -\frac{1}{2}\rho\kappa\diff\rho\wedge \calB~,
\eea
where to simplify resulting equations it is useful to define the function
\bea
\kappa &\equiv & \frac{\ex^{-6\Delta}}{1-\|\xi\|^2}~.
\eea
Decomposing as outlined above, this becomes
\bea\label{SUSY1dec}
\partial_\tau \calBhat - \hatd \calB_\tau &=& 0~,\nonumber\\
\partial_\rho \calB &=& -\frac{1}{2}\rho\kappa\calB~,\nonumber\\
\hatd \calBhat &=& 2\Jhat_3~.
\eea
Note here that everything is invariant under $\partial_\psi$. The integrability condition for 
(\ref{SUSY1}) immediately implies that $\partial_\tau \Jhat_3=0=\hatd \Jhat_3$, while combining the component 
\bea
\hatd\left(\kappa\calB_\tau\right) - \partial_\tau \left(\kappa\calBhat\right) &=& 0~,
\eea
with the 
first and last equation in (\ref{SUSY1dec}) leads to the conclusion
\bea
\partial_\tau\kappa &=& 0 \ = \ \hatd  \kappa~.
\eea
Given (\ref{norm}), this then implies
\bea
\partial_\tau \Delta &=& 0 \ = \  \hatd \Delta~,
\eea
so that the warp factor $\Delta$, and the related functions $\kappa$, $\zeta$, $|S|$ and $\|\xi\|$, all depend 
only on the coordinate $\rho$! 

The other two equations in (\ref{SUSYs}) may be analyzed similarly. Rather than present all the details, which 
are straightforward but rather long, we simply present the final result. Defining
$\Omegahat=\Jhat_1+\ii\Jhat_2$, the supersymmetry conditions (\ref{SUSYs}) are equivalent to the equations
\begin{equation}\label{summary}
\boxed{
\begin{array}{rcl}
\partial_\rho \calB & = &  -\frac{1}{2}\rho\kappa\calB~,
\qquad \! \!\! \!\big[\partial_\rho \Omegahat\big]_+ \  = \  - \frac{1}{2}\rho\kappa\Omegahat~, \\[2mm]
\hatd\calBhat & = & 2\Jhat_3~,\qquad 
~~~\big[\partial_\tau \Omegahat\big]_+ \ = \ -\ii \hh \Omegahat~, \quad   \quad \hatd\Omegahat \ = \ \big( \big[\partial_\tau\Omegahat\big]_--\ii \hh\Omegahat \big)\wedge
 \frac{\calBhat}{\calB_\tau}~,\\[2mm] 
\partial_\tau \calBhat & = & \hatd\calB_\tau ~, \qquad \, ~~\big[\partial_\tau\Omegahat\big]_- \ = \ \zeta\calB_\tau\left(\big[\rho\partial_\rho\Jhat_2\big]_--
\frac{\ii}{\|\xi\|^2}\big[\rho\partial_\rho\Jhat_1\big]_-\right)~.
\end{array}
}
\end{equation}
Here we have defined the function
\bea
\hh & \equiv & \zeta\calB_\tau\left(\frac{1}{2}\rho\partial_\rho\log \kappa - \rho^2\kappa\right)~,
\eea
and the notation $\big[\cdot\big]_\pm$ denotes the self-dual and anti-self-dual parts 
of a two-form along the four-dimensional $SU(2)$-structure space. In particular, of course 
$\Jhat_I$, $I=1,2,3$, form a basis for the self-dual forms.
We also note that the integrability condition for the three equations in the first column of (\ref{summary}) gives
\bea
\partial_\tau \Jhat_3 &  = &  0~, \qquad \partial_\rho \Jhat_3 \ = \  - \frac{1}{2}\rho\kappa\Jhat_3~, \qquad \hatd \Jhat_3 \ =\  0~.
\eea

As an aside comment, we notice that a subset of the equations in (\ref{summary}) may be re-interpreted as equations for a dynamical 
\emph{contact-hypo} structure on a five-dimensional space \cite{Conti,bedulli}.  Here we decompose the seven-dimensional manifold
under a $1+1+5$ split, where the two transverse directions are parametrized by the coordinates $\rho$ and $\psi$. The 
$({\cal B},J_I)$ then define a contact-hypo structure (at fixed $\rho$) obeying the equations 
\bea
\tilde{\diff} \mathcal{B} \ = \  2\hat J_3 ~, \qquad \quad \tilde{\diff} \Omegahat \ = \ 
\big( \big[\partial_\tau\Omegahat\big]_--\ii \hh\Omegahat \big)\wedge \frac{\calB}{\calB_\tau}~,
\eea
where $\tilde{\diff} \equiv \diff\tau\wedge \frac{\partial}{\partial\tau}  + \hatd$. Note that when $\big[\partial_\tau\Omegahat\big]_-=0$ these become
the conditions characterizing a Sasaki-Einstein five-manifold. However, in this paper we will not pursue further this point of view. 
 
We emphasize again that since $\Delta$ is a  function only of $\rho$, this implies that the derived functions
$\kappa$, $\zeta$, and $\|\xi\|$ also depend only on $\rho$. We conclude by writing an even more explicit
expression for the flux given in (\ref{flux}):
\begin{equation}
\boxed{
\begin{array}{rcl}
F &=& \frac{1}{\|\xi\|}\left(12\ex^{6\Delta}\|\xi\|^2\partial_\rho \Delta - 6\rho\right)E_{12}\wedge J_1 
-12 \ex^{6\Delta}\partial_\rho\Delta E_{13}\wedge J_2 \\
&&- m\frac{\sqrt{1-\|\xi\|^2}}{\|\xi\|}E_{23}\wedge J_1 + \frac{m}{6}\ex^{3\Delta}(1-\|\xi\|^2)E_{13}\wedge \big[\partial_\rho \hat{J}_2\big]_-\\
&&+ \frac{m}{6}\ex^{3\Delta}\frac{(1-\|\xi\|^2)}{\|\xi\|}E_{12}\wedge \big[\partial_\rho \hat{J}_1\big]_-~.
\end{array}}
\end{equation}
This expression is particularly useful for proving sufficiency of the differential system in section \ref{sec:NS}.

We shall investigate the general equations (\ref{summary}), in a special case, in section \ref{sec:tauKilling}, reducing them 
to a single second order ODE in $\rho$.

\section{M2-brane solutions}\label{sec:contact}

In this section we further elaborate on the geometry and physics of solutions with $m\neq0$. 
In particular we show that all such solutions admit a canonical contact structure, 
for which the R-symmetry Killing vector $\xi$ is the Reeb vector field. Many physical 
properties of the solutions, such as the free energy and scaling dimensions of 
BPS wrapped M5-branes, can be expressed purely in terms of this contact structure. 
This section is essentially an expansion of the material in \cite{Gabella:2011sg}, as advertized 
in that reference.

\subsection{Contact structure}

When $m\neq 0$ we may define a one-form $\sigma$ via
\bea
P &\equiv & \zeta \sigma~,
\eea
where $P$ is the one-form bilinear defined in the 
second line in (\ref{vectors}). In terms of our frame (\ref{frame}), we then have
\bea
\sigma &=& \frac{1}{\|\xi\|}E_1 + \frac{|S|\sqrt{1-\|\xi\|^2}}{\zeta \|\xi\|}E_3~, \nonumber\\[2mm]
 & = & \frac{1}{4} \left[ \diff \psi + {\cal A} + \left(\tfrac{6}{m}\right)^2 \rho^2 (\diff \tau + {\cal A})\right]~. 
\eea
Up to a factor of $m/6$, the one-form inside the square bracket on the left hand side of the first equation in (\ref{SUSYs}) is in fact $\sigma$. 
Thus we read off
\bea
\diff\sigma &=& \frac{12}{m}\ex^{3\Delta}\left(J_3 - \|\xi\|E_2\wedge E_3\right)~,
\eea
%\bea
%\diff \sigma &=& -\frac{12}{m}\ex^{3\Delta}\, \Real \bar\chi_+^c\gamma_{(2)}\chi_-~,
%\eea
and a simple algebraic computation then leads to
\bea\label{contactvolume}
\sigma\wedge (\diff\sigma)^3 &=& \frac{2^73^4}{m^3}\ex^{9\Delta}\vol_7~.\eea
Here
\bea
\vol_7 &\equiv & -E_1\wedge E_2\wedge E_3\wedge \vol_4 \ = \ -\frac{1}{2}E_1\wedge E_2\wedge E_3\wedge J_3\wedge J_3~,
\eea
denotes the Riemannian volume form of $Y_7$ (with a convenient choice of orientation). 
It follows that when $m\neq 0$, the seven-form $\sigma\wedge (\diff\sigma)^3$ is a nowhere-zero 
top degree form on $Y_7$, and thus by definition $\sigma$ is a contact form on $Y_7$.

Again, straightforward algebraic computations using the Fierz identity in appendix \ref{app:useful} lead to 
\bea
\xi\lrcorner\sigma \ = \ 1~, \qquad  \xi\lrcorner\diff\sigma \ = \ 0~.
\eea
This implies that the Killing vector field $\xi$ is also the unique Reeb vector field 
for the contact structure defined by $\sigma$.

\subsection{Flux quantization}

When $m\neq 0$, equation (\ref{mF}) immediately leads to the natural gauge choice 
\bea
F &=& \diff A~,
\eea
where $A$ is the global three-form
\bea\label{A}
A & \equiv & \frac{6}{m}\ex^{6\Delta} \mathrm{Im}\, \bar{\chi}_+^c \gamma_{(3)}\chi_-~.
\eea
In terms of our frame, this reads
\bea
A&=& \frac{6}{m}\ex^{6\Delta}\Big[|S|J_2\wedge E_2 - \frac{1}{\|\xi\|}J_1\wedge\left(|S|E_3 + \frac{m}{6}\ex^{-3\Delta}\sqrt{1-\|\xi\|^2}E_1\right)\Big]~.
\eea
Notice that, either using the last expression or using (\ref{charge2}), we find that 
\bea
\mathcal{L}_\xi A &=& 0~.
\eea
Of course, one is free to add to $A$ any closed three-form $a$, which will result in the same curvature 
$F$
\bea\label{gaugeA}
A & \rightarrow & A + \frac{1}{(2\pi \ell_p)^3}a~.
\eea
If $a$ is exact this is a gauge transformation of $A$ and leads to a physically equivalent
M-theory background. In fact more generally if $a$ has integer periods then the transformation
(\ref{gaugeA}) is a large gauge transformation of $A$, again leading to an equivalent
solution. It follows that only the cohomology class of $a$ in the torus $H^3(Y_7;\R)/H^3(Y_7;\Z)$
is a physically meaningful parameter, and this corresponds to a marginal parameter in
the dual CFT. In fact the free energy will be independent of this choice of $a$, which is
why we have set $a = 0$ in (\ref{A}). There is also the possibility of adding discrete torsion
to $A$ when $H^4_{\mathrm{torsion}}(Y_7;\Z)$ is non-trivial, but we will not discuss this here.

The flux quantization condition in eleven dimensions is 
\bea\label{flux11}
N &=& -\frac{1}{(2\pi\ell_p)^6}\int_{Y_7} *_{11} G+ \frac{1}{2}C\wedge G~,
\eea
where $N$ is the total M2-brane charge. Dirac quantization requires that $N$ is an integer. 
Substituting our ansatz (\ref{ansatz}) into (\ref{flux11}) leads to
\bea
N &=& \frac{1}{(2\pi\ell_p)^6}\int_{Y_7} m \ex^{3\Delta}\vol_7 - \frac{1}{2}A\wedge F~,
\eea
where $\vol_7$ denotes the Riemannian volume form for $Y_7$.
By far the 
simplest way to evaluate $A\wedge F$ is to use the identity (\ref{useful1}) with 
$\mathcal{C}=1$. Using (\ref{A}), this immediately leads to an expression for 
$A\wedge F$ in terms of $\vol_7$, and using (\ref{contactvolume})
we obtain
\bea\label{Nexplicit}
N &=& \frac{1}{(2\pi\ell_p)^6}\frac{m^2}{2^53^2}\int_{Y_7}\sigma\wedge (\diff \sigma)^3~.
\eea
In particular, we see that $m\neq 0$ leads to a non-zero  M2-brane charge $N$.

\subsection{The free energy}

The effective four-dimensional Newton constant $G_4$ is computed by dimensional 
reduction of eleven-dimensional supergravity on $Y_7$. More precisely, by definition $1/16\pi G_4$ is 
the coefficient of the four-dimensional Einstein-Hilbert term, in Einstein frame. A standard 
 computation leads to the formula
\bea\label{newton}
\frac{1}{16\pi G_4} &=& \frac{\pi\int_{Y_7} \ex^{9\Delta}\vol_7}{2(2\pi\ell_p)^9}~.
\eea
On the other hand, $G_4$ also determines the gravitational free energy $\mathcal{F}_{\mathrm{AdS}}$ 
\bea\label{F}
\mathcal{F}_{\mathrm{AdS}}&\equiv & - \log |Z| \ = \ \frac{\pi}{2G_4}~.
\eea
Here the left hand side of (\ref{F}) is the free energy of the unit radius AdS$_4$ computed in Euclidean quantum gravity, 
where $Z$ is the gravitational partition function. Thus in the supergravity approximation, $\mathcal{F}_{\mathrm{AdS}}$ is simply
the four-dimensional on-shell Einstein-Hilbert action, which has been regularized to give the finite result on 
the right hand side of (\ref{F}) using the boundary counterterm subtraction method of \cite{Emparan:1999pm}. 
Via the AdS/CFT correspondence, $\mathcal{F}_{\mathrm{AdS}}=\mathcal{F}_{\mathrm{CFT}}\equiv\mathcal{F}$, where 
$\mathcal{F}_{\mathrm{CFT}}$ is the free energy of the dual CFT on the conformal boundary $S^3$ of AdS$_4$.
Combining (\ref{newton}) and (\ref{F}) then leads to the supergravity formula
\bea\label{merlin}
\mathcal{F}&=& \frac{4\pi^3\int_{Y_7} \ex^{9\Delta}\vol_7}{(2\pi\ell_p)^9}~.
\eea
Combining (\ref{Nexplicit}), (\ref{merlin}) and (\ref{contactvolume}) leads
to our final formula
\begin{equation}\label{freeenergy}
\boxed{
\mathcal{F} \ = \ N^{3/2} \sqrt{\frac{32\pi^6}{9\int_{Y_7} \sigma\wedge (\diff\sigma)^3}}~.
}
\end{equation}
We see that the famous $N^{3/2}$ scaling behaviour of the free energy of $N$ M2-branes 
continues to hold in the most general $\mathcal{N}=2$ supersymmetric case with flux turned on. Moreover, the coefficient 
is expressed purely in terms of the contact volume of $Y_7$. In the Sasaki-Einstein case this 
agrees with the Riemannian volume computed using $\vol_7$, but more generally the two volumes 
are different. 
The contact volume has the property, in the sense described precisely in 
appendix B of \cite{Gabella:2010cy}, that it depends only on the Reeb vector field $\xi$ 
determined by the contact structure. In particular, if we formally consider 
varying the contact structure of a given solution, the contact volume is a strictly 
convex function of the Reeb vector field $\xi$. It is of course natural to conjecture 
that this function is related as in (\ref{F}) to minus the logarithm of the field theoretic $|Z|$-function defined in 
\cite{Jafferis:2010un}, as a function of a trial R-symmetry in the dual supersymmetric field theory 
on $S^3$. This was conjectured in the Sasaki-Einstein case in \cite{Martelli:2011qj}, and has by 
now been verified in a large number of examples, including infinite families \cite{Amariti:2012tj}. 
The contact volume has the desirable property that it can be computed using topological and 
fixed point theorem methods, so that one can compute the free energy of a solution essentially 
knowing only its Reeb vector field. We will illustrate this with the class of solutions in section \ref{sec:tauKilling}.

Finally, the scaling symmetry of eleven-dimensional supergravity in which the metric $g_{11}$ and 
four-form $G$ have weights two and three, respectively, leads to a symmetry in which 
one shifts $\Delta\rightarrow \Delta+c$ and simultaneously scales $m\rightarrow \ex^{3c}m$, 
$F\rightarrow \ex^{3c}F$, where $c$ is any real constant. We may then take the metric on 
$Y_7$ to be of order $\mathcal{O}(N^0)$, and conclude from the quantization condition 
(\ref{flux11}), which has weight 6 on the right hand side,  that $\ex^\Delta=\mathcal{O}(N^{1/6})$. 
It follows that the AdS$_4$ radius, while dependent on $Y_7$, is $R_{\mathrm{AdS}_4}=\ex^\Delta=\mathcal{O}(N^{1/6})$, 
and that the supergravity approximation we have been using is valid only in the $N\rightarrow\infty$ limit.

\subsection{Scaling dimensions of BPS wrapped M5-branes}\label{sec:M5}

A probe M5-brane whose world-space is wrapped on a generalized calibrated five-submanifold $\Sigma_5 \subset Y_7$ and which moves along a geodesic in AdS$_4$ is expected to correspond to 
a BPS operator  $\mathcal{O}_{\Sigma_5}$
in the dual three-dimensional SCFT.  
In particular, when $Y_7$ is a Sasaki-Einstein manifold, the scaling dimension of this operator can be calculated from the volume of the 
five-submanifold $\Sigma_5$ \cite{Berenstein:2002ke}. In this section 
we show that a simple generalization of this correspondence holds for the general ${\cal N}=2$ supersymmetric 
AdS$_4\times Y_7$ 
solutions treated in this paper.\footnote{Such supersymmetric M5-branes exist only for certain boundary conditions 
\cite{Klebanov:2010tj,Benishti:2010jn}, and  our discussion
 here applies to these cases.} 

Given a Killing spinor $\epsilon$ of eleven-dimensional supergravity, it is simple to derive the following BPS bound for the M5-brane \cite{Barwald:1999ux,Martelli:2003ki} 
\bea
\epsilon^\dagger  \epsilon \, L_\text{DBI}\, \vol_5 &  \geq  &  \left[\frac{1}{2} (j^*{k}\lrcorner  H)\wedge H +  j^*{\mu} \wedge H   +  j^*{\nu}\right]~. \label{mainbound}
\eea
Here $H$ is the three-form on the M5-brane, defined by $H= h + j^*C$ where $h$ is closed and $j^*$ denotes the pull-back to the M5-brane world-volume. 
The one-form ${k}$, two-form ${\mu}$ and  five-form ${\nu}$ are defined \cite{Gauntlett:2002fz} by the eleven-dimensional bilinears  
\bea
k &\equiv &  \bar \epsilon \Gamma_{(1)}\epsilon~,\qquad 
 \mu  \ \equiv \    \bar \epsilon \Gamma_{(2)}\epsilon~, \qquad \nu  \ \equiv \    \bar \epsilon \Gamma_{(5)}\epsilon~,\eea
and  $\vol_5$ is the volume form on the world-space of the M5-brane.
We have defined $\bar \epsilon \equiv \epsilon^\dagger \Gamma_0$ as usual. 

The bound (\ref{mainbound}) follows from the inequality
\bea
\|\mathcal{P}_- \epsilon\|^2 & = & \epsilon^\dagger \mathcal{P}_- \epsilon \geq 0~, 
\eea
where $\proj \equiv(1 - \tilde \Gamma)/2$ is the $\kappa$-symmetry projector and $\tilde \Gamma$ is the traceless Hermitian product structure
\bea
\tilde{\Gamma} &\equiv& 
\frac{1}{L_{\text{DBI}}} \Gamma_{0} 
\left[ \frac{1}{4}  (j^* \Gamma)^a (H^*\lrcorner H)_a  + \frac{1}{2!} (j^* \Gamma)^{a_1a_2}  H^*_{a_1a_2} 
+ \frac{1}{5!} (j^*\Gamma)^{a_1\cdots a_5} \varepsilon_{a_1\cdots a_5} \right] ~.
\eea
Here $a,a_1\dots a_5 = 1,\dots, 5$, where the two-form $H^* \equiv *_5 H$ is the world-space dual of $H$. This bound is saturated if and only if $\mathcal{P}_- \epsilon = 0$ 
and corresponds to a probe M5-brane preserving supersymmetry. 

We write the AdS$_4$ metric in global coordinates (\emph{cf.} footnote \ref{footing}) and choose
the static gauge embedding  $\{ \ttt = \sigma^0 , x^m = \sigma^m \}$,  where  $\ttt$ is global
time in AdS$_4$ and  $x^m$, with $m=1,\dots, 5$, are coordinates on $Y_7$. The Dirac-Born-Infeld 
Lagrangian $L_\text{DBI}$ is then defined by $L_{\text{DBI}} = \sqrt{\det(\delta_m^{\ n} + H_m^{* n})}$.
The vector $k_\sharp$ dual to the one-form $k$ is a time-like Killing vector, which using the explicit form of the 
eleven-dimensional ${\cal N}=2$ Killing spinor (\ref{n2spinoransatz}), and an appropriate choice of AdS$_4$ spinors $\psi_i$, reads
\bea\label{dualk}
k_\sharp & = & \partial_t + \frac{1}{2} \xi ~.
\eea
Accordingly, $\epsilon^\dagger \epsilon = k_\sharp^0 = \tfrac{1}{2} \ex^\Delta\cosh\varrho$, and hence the bound (\ref{mainbound}) 
is saturated when $\varrho=0$
({\it i.e.} the M5-brane is at the centre of AdS$_4$) and 
\bea
\frac{\me^{\Delta}}{2} L_\text{DBI}\, \vol_5 &  = & \left[\frac{1}{2} ( j^*k\lrcorner H)\wedge H +   j^*\mu \wedge H   +   j^*\nu\right]~.\label{satur}
\eea

The energy density of an M5-brane can be computed by solving the  Hamiltonian constraints \cite{Barwald:1999ux,Martelli:2003ki}. 
For the static gauge embedding and $\varrho = 0$ these lead to 
 \bea
 {\cal E} & =&  P_\ttt \ =\ T_{\text{M5}} \left(\frac{\ex^\Delta  }{2} L_{\text{DBI}} + \mathcal{C}_\ttt\right)~,
 \eea 
 where  $T_{\text{M5}}=2\pi/(2\pi \ell_p)^6$ is the M5-brane tension and the contribution from the Wess-Zumino coupling is $
{\cal C}_\ttt \vol_5  = \de_\ttt \lrcorner C_6 - \tfrac{1}{2} (\de_\ttt \lrcorner C) \wedge (C- 2H)$,
with the potential $C_6$ defined through $\diff C_6 = *_{11}G +\tfrac{1}{2}C\wedge G$. However, from the explicit expression of
 $C$ one can check that we have ${\cal C}_\ttt=0$.
The M5-brane energy is then given by 
\bea
E_\text{M5}& =&  T_{\text{M5}} \int_{\Sigma_5} \frac{\ex^\Delta  }{2} L_{\text{DBI}}\, \vol_5 \, =\, 
T_{\text{M5}}  \int_{\Sigma_5} \frac{1}{4} (\xi \lrcorner H)\wedge H + j^*\mu \wedge H   +   j^*\nu ~,
\label{trivial}
\eea
where we used \eqref{dualk}.
Let us briefly discuss this expression for the energy.   With our gauge choice (\ref{A}) for the three-form potential, 
in general we have $H=A+h$, where $h$ is a closed three-form.  If $h$ is exact and 
invariant\footnote{One should obviously require that $\de_\ttt$ and $\xi$ generate symmetries of the M5-brane action.} 
under $k_\sharp$, namely $h=\diff b$ with ${\cal L}_{k_\sharp} b=0$, then one can check that the integral does not depend on $h$. To see this, 
one has to recall that ${\cal L}_{k_\sharp} A=0$, use the results of
\cite{Gauntlett:2002fz}, and  apply Stokes' theorem repeatedly. If $h$ is not exact, \emph{a priori} it will contribute to the energy, 
and hence we expect the dimension of the dual operator to be affected. We  leave an investigation of this interesting possibility for future work, 
and henceforth set $H=A$. In particular, $A$ is expressed  as a bilinear of $\chi_\pm$ in \eqref{A}.

Using the explicit form of the eleven-dimensional ${\cal N}=2$ Killing spinor (\ref{n2spinoransatz}) and the static gauge embedding one derives
\bea
\iota^*k &=& \tfrac{1}{2}  \ex^{2\Delta} K ~,\nn \\
\iota^*\mu &=& 4 \ex^{3\Delta} \left\{ -\tfrac{1}{8} \Imag[ \bar \chi_+^c \gamma_{(2)} \chi_- ] 
+ \mathrm{Im}[\bar\psi^+_1(\psi^+_2)^c] \Real[ \bar \chi_+^c \gamma_{(2)} \chi_- ] \right \}~,\nn \\
\iota^*\nu  &=& 4 \ex^{6\Delta} \star \left\{ \tfrac{1}{8} \Real[ \bar \chi_+^c \gamma_{(2)} \chi_- ] +  \mathrm{Im}[\bar\psi^+_1(\psi^+_2)^c]\Imag[ \bar \chi_+^c \gamma_{(2)} \chi_- ] \right\}~,
\eea
where $\iota^*$ denotes a pull-back to $Y_7$, and where the constant scalar bilinear $\mathrm{Re}[\bar\psi^+_1(\psi^+_2)^c]$ is rescaled for convenience to $\frac{1}{8}$. The $\chi_\pm$ bilinears can then be expressed in terms of $E_i$ and $J_I$. The non-constant scalar $\mathrm{Im}[\bar\psi^+_1(\psi^+_2)^c]$ drops 
out of the calculation and one arrives at\footnote{The sign arises from our choice of conventions, {\it cf.} \cite{Gabella:2009wu}.} 
\bea
\frac{1}{2} (j^*{k}\lrcorner  H)\wedge H +  j^*{\mu} \wedge H   +  j^*{\nu} = - \frac{m^2}{2^6 3^2}  \sigma \wedge (\diff \sigma)^2 ~.
\eea

Hence we get the remarkably simple result
\bea
E_\text{M5}  &=&  - T_{\text{M5}}\frac{m^2}{2^6 3^2} \int_{\Sigma_5} \sigma \wedge (\diff \sigma )^2 ~.
\eea
Combining the latter with (\ref{Nexplicit}), and 
identifying $\Delta(\mathcal{O}_{\Sigma_5})$ with 
the energy  $E_\text{M5}$ in global AdS,
leads straightforwardly to the formula
\begin{equation}
\label{Delta}
\boxed{
\Delta(\mathcal{O}_{\Sigma_5}) \ = \ \pi N \left|\frac{\int_{\Sigma_5}\sigma\wedge (\diff\sigma)^2}{
\int_{Y_7} \sigma\wedge (\diff\sigma)^3} \right|~.
}
\end{equation}
The scaling dimensions of operators dual to BPS wrapped M5-branes are thus also 
determined purely by the contact structure. As for the contact volume of 
$Y_7$, the right hand side of (\ref{Delta}) can again be computed from a knowledge 
of $\Sigma_5$ and the Reeb vector field $\xi$.

\section{Special class of solutions: $\partial_\tau$ Killing}\label{sec:tauKilling}

Since the general system of supersymmetry equations presented in section 
\ref{sec:reduction} is rather complicated, in this section 
we impose a single simplifying assumption, namely that 
$\partial_\tau$ is a \emph{Killing vector} field for the metric\footnote{Note that we are not requiring that 
$\de_\tau$ generates  a symmetry of the full solution. Indeed we will show that in general 
the flux $F$ is \emph{not} invariant under $\de_\tau$.} $g_7$ . 
There are two motivations 
for this. Firstly, it is clearly a natural geometric condition. Secondly, 
the only solution in the literature in the $m\neq 0$ class 
that is not Sasaki-Einstein is the Corrado-Pilch-Warner solution \cite{Corrado:2001nv}. 
This solution describes the infrared fixed point of a massive deformation 
of the maximally supersymmetric AdS$_4\times S^7$ solution, and has the same topology 
but with non-standard metric on $S^7$ and flux. We will first show that 
the assumption that $\partial_\tau$ is Killing immediately leads 
to the four-metric $g_{SU(2)}$ being conformal to a K\"ahler-Einstein metric, and that the supersymmetry conditions then 
entirely reduce to a single second order non-linear ODE. The Corrado-Pilch-Warner 
solution is a particular solution to this ODE, with $g_{SU(2)}$ being (conformal to) the standard Fubini-Study metric on $\mathbb{CP}^2$. We will then show numerically that there exists 
a second solution, dual to the infrared fixed point of a cubic deformation 
of $N$ M2-branes at a general $\mathrm{CY}_3\times \C$ singularity, where 
$\mathrm{CY}_3$ denotes any Calabi-Yau three-fold cone.
In particular, when $\mathrm{CY}_3=\C^3$ endowed with a flat metric, 
this leads to a new, smooth $\mathcal{N}=2$ supersymmetric AdS$_4\times S^7$ solution.

\subsection{Further reduction of the equations}

Let us analyze the conditions (\ref{summary}), with the assumption that $\partial_\tau$ is Killing. 
Notice that the latter implies
\bea
\big[\partial_\tau \Jhat_I\big]_\pm &= & \partial_\tau \big[\Jhat_I\big]_\pm \ = \ \begin{cases} \ \partial_\tau \Jhat_I \\ \ 0\end{cases}~.
\eea
The left hand side of last equation in (\ref{summary}) is thus identically zero. Taking the real and imaginary parts of the right hand 
side then implies that $\partial_\rho \Omegahat$ is self-dual. The plus subscripts may then be dropped in the second 
line of (\ref{summary}), and we see that
\bea
\partial_\rho \Jhat_I &=& -\frac{1}{2}\rho\kappa \Jhat_I~,
\eea
holds for all $I=1,2,3$. Recalling that $\kappa$ is always a function only of $\rho$, 
we may introduce the rescaled $SU(2)$ structure 
\bea
\Jhat_I &\equiv & f(\rho) \bbJ_I~, \qquad I=1,2,3~,
\eea
and see that provided $f(\rho)$ satisfies the differential equation
\bea\label{diff1rho}
\frac{\diff f}{\diff\rho} &=& -\frac{1}{2}\rho\kappa f~,
\eea
then the $SU(2)$-structure two-forms $\bbJ_I$ are independent of $\rho$. 

Similarly, the Killing condition on $\partial_\tau$ implies that $\calB_\tau$ and $\calBhat$ are independent of $\tau$, 
and the first equation in (\ref{summary}) then implies that $\calB_\tau=\calB_\tau(\rho)$ depends only on $\rho$. 
We may then similarly solve the second equation in (\ref{summary}) by rescaling
\bea
\calB &\equiv & f(\rho)\bbB~,
\eea
and deduce that $\bbB$ is independent of both $\tau$ and $\rho$. Similarly writing
\bea
\bbB &\equiv & \bbB_\tau\diff\tau + \hat{\bbB}~,
\eea
where now $\bbB_\tau$ is a constant,
the remaining equations in (\ref{summary}) are
\bea\label{remaining}
\hatd \hat{\bbB} &=& 2\bbJ_3~, \qquad \hatd (\bbJ_1+\ii \bbJ_2) \ = \ -\ii f\zeta \left(\frac{1}{2}\rho\partial_\rho\log \kappa - \rho^2\kappa\right)(\bbJ_1+\ii \bbJ_2)\wedge \hat{\bbB}~,\nonumber\\
\partial_\tau(\bbJ_1+\ii \bbJ_2) &=& - \ii u (\bbJ_1+\ii \bbJ_2)~.
\eea
Since the blackboard script quantities are independent of $\rho$, the second equation in (\ref{remaining}) implies that
\bea\label{diff2rho}
f\zeta \left(\frac{1}{2}\rho\partial_\rho\log \kappa - \rho^2\kappa\right) &=& -\gamma~,
\eea
which is {\it a priori} a function of $\rho$, is in fact a constant. At this point we should recall the definitions 
\bea
\zeta &=& \frac{m}{6}\ex^{-3\Delta}~, \qquad \kappa \ = \ \frac{\ex^{-6\Delta}}{1-\ex^{-6\Delta}\left[\left(\frac{m}{6}\right)^2+\rho^2\right]}~.
\eea
In order to remove the explicit factors of $m$, and write everything in terms of a single function, it is convenient to rescale 
\bea
r & \equiv & \frac{6}{m}\rho~, \qquad \alpha^2(r) \ \equiv \ \left(\frac{m}{6}\right)^2\kappa~.
\eea
In terms of these new variables, the differential equations (\ref{diff1rho}), (\ref{diff2rho}) read
\begin{equation}\label{ODEs}
\boxed{\begin{array}{rcl}
f' &=& -\frac{1}{2}r\alpha^2f~,\\[1.5mm]
\frac{\left(r\alpha' - r^2\alpha^3\right)f}{\sqrt{1+(1+r^2)\alpha^2}} &=& -\gamma~,
\end{array}}
\end{equation}
which are a coupled set of first order ODEs for the functions $f(r)$, $\alpha(r)$, and from 
henceforth a prime will denote derivative with respect to the coordinate $r$. The remaining supersymmetry conditions 
(\ref{remaining}) now simplify to
\bea
\hatd \hat{\bbB} &=& 2\bbJ_3~, \qquad \hatd (\bbJ_1+\ii \bbJ_2) \ = \ \ii \gamma (\bbJ_1+\ii \bbJ_2)\wedge \hat{\bbB}~,\nonumber\\
\partial_\tau(\bbJ_1+\ii \bbJ_2) &=&  \ii \gamma {\bbB}_\tau(\bbJ_1+\ii \bbJ_2)~.
\eea
Here both $\gamma$ and $\bbB_\tau$ are constants. The first line says that 
the four-metric defined by $(\bbJ_1,\bbJ_2,\bbJ_3)$ is K\"ahler-Einstein 
with Ricci tensor satisfying $\mathrm{Ric} = 2\gamma g_{\mathrm{KE}}$. 
The second equation is solved simply by multiplying $\bbJ_1+\ii\bbJ_2$ by 
a phase $\ex^{-\ii\gamma\bbB_\tau}$, so that 
everything is independent of $\tau$. 

To conclude, given any K\"ahler-Einstein four-metric $g_{\mathrm{KE}}$ with Ricci curvature $\mathrm{Ric}=2\gamma g_{\mathrm{KE}}$, 
a solution to the ODE system (\ref{ODEs}) leads to a (local) supersymmetric AdS$_4$ solution 
with internal seven-metric being
\bea\label{7dmetric}
g_7 & = & \frac{f\alpha}{4\sqrt{1+(1+r^2)\alpha^2}}g_{\mathrm{KE}} + 
\frac{\alpha^2}{16}\Bigg[\diff r^2 + \frac{ r^2 f^2}{1+r^2}(\diff\tau + {A}_\mathrm{KE})^2\nonumber\\[1.8mm]
&&+ \frac{1+r^2}{1+(1+r^2)\alpha^2}\left(\diff\psi + \frac{f}{1+r^2}(\diff\tau+{A}_\mathrm{KE})\right)^2\Bigg]~,
\eea
and flux
\bea
F &=& \frac{m^2\ex^{-3\Delta}\alpha}{3^3\cdot 2^7} \left(\gamma m \ex^{-3\Delta}\alpha(1+r^2) - 9 r^2 f\right) (\diff\psi -\diff\tau) \wedge \frac{\diff r}{r} \wedge \mathbb{J}_1 \nonumber\\[1.8mm]
&+&\frac{\gamma m^3\ex^{-6\Delta}\alpha^2f}{3^3\cdot 2^7}  (\diff\tau+ A_{\mathrm{KE}}) \wedge \Big(\frac{\diff r}{r} \wedge \mathbb J_1 + (\diff\psi -\diff\tau) \wedge \mathbb J_2 \Big) ~,
\label{newf}
\eea
where we have written the latter in terms of three functions $f, \alpha,\ex^\Delta$ in order to simplify the expression slightly. 
However, recall that the warp factor is related to $\alpha$ via
\bea
\ex^{6\Delta} & = & \left(\frac{m}{6}\right)^2 \left( 1 + r^2 + \alpha^{-2}\right)~.
\eea
Here we have denoted $A_\mathrm{KE}\equiv\hat{\bbB}$, and without loss of generality 
we have set $\bbB_\tau=1$ by rescaling the $\tau$ coordinate. 
From (\ref{newf}) we see explicitly that ${\cal L}_{\de_\tau} F\neq 0$,  since the holomorphic two-form on the K\"ahler-Einstein base
satisfies  ${\cal L}_{\de_\tau} (\mathbb{J}_1 + \ii \mathbb{J}_2) =  \ii \gamma(\mathbb{J}_1 + \ii \mathbb{J}_2)$. Therefore,
 as anticipated at the beginning of this section, $\de_\tau$ does not generate  a symmetry of the full solution.
If $\gamma>0$ then by rescaling $f$ we may also without loss of generality set $\gamma=3$.
The local one-form $\gamma A_{\mathrm{KE}}$ is globally a connection on the anti-canonical bundle of the 
K\"ahler-Einstein four-space. 
Notice that we may algebraically eliminate $\alpha(r)$ from the first equation in (\ref{ODEs}) to obtain 
the single second order ODE for $f(r)$
\bea\label{2ndorder}
3rf'^2+f(rf''-f')&=&\gamma\sqrt{-2f'\left[rf-2(1+r^2)f'\right]}~.
\eea

\subsection{The Corrado-Pilch-Warner solution}\label{sec:CPW}

We begin by noting that the following is an explicit solution to the ODE system (\ref{ODEs})
\bea\label{CPW}
f(r)&=& \gamma\left(2-\frac{r}{\sqrt{2}}\right)~,\qquad
\alpha(r) \ =\  \sqrt{\frac{2}{r(2\sqrt{2}-r)}}~.
\eea
Taking the K\"ahler-Einstein metric to be simply the standard Fubini-Study metric on $\mathbb{CP}^2$, 
and with $r\in [0,2\sqrt{2}]$, we claim this is precisely the AdS$_4\times S^7$ solution described in 
\cite{Corrado:2001nv}. In fact the authors of \cite{Corrado:2001nv} conjectured 
that one should be able to replace $\mathbb{CP}^2$ by any other K\"ahler-Einstein 
metric (with positive Ricci curvature) to obtain another supergravity solution. 
This was shown in \cite{Ahn:2009sk} for the special case in which one uses 
the K\"ahler-Einstein metrics associated to the $L^{abc}$ Sasaki-Einstein manifolds \cite{Cvetic:2005ft}, \cite{Martelli:2005wy}. 
We can immediately read off the warp factor
\bea\label{PWzeta}
\frac{m}{6}\ex^{-3\Delta} \ = \ \zeta &=& \frac{\alpha}{\sqrt{1+(1+r^2)\alpha^2}} \ = \ \frac{1}{1+\frac{r}{\sqrt{2}}}~.
\eea
Comparing our $\diff r^2$ component of the metric (\ref{7dmetric}) to the $\diff \mu^2$ component 
of the metric in \cite{Ahn:2009sk}, we are led to the identification
\bea
r &=& 2\sqrt{2}\sin^2\mu~.
\eea
It is then straightforward to see that our metric (\ref{7dmetric}) coincides with the metric in 
\cite{Ahn:2009sk}, and using (\ref{newf}) also that the fluxes agree.

\subsection{Deformations of $\mathrm{CY}_3\times\C$ backgrounds}\label{sec:def}

The Corrado-Pilch-Warner solution fits into a more general class of solutions 
obtained by deforming the theory on $N$ M2-branes at the conical singularity of 
the Calabi-Yau four-fold $\mathrm{CY}_3\times\C$. In this section 
we give a unified treatment, in particular recovering the field theory result
in \cite{Jafferis:2011zi} for the free energy of such theories using our contact volume formula (\ref{freeenergy}).

We begin by 
taking $g_{\mathrm{KE}}$ to be the (local) K\"ahler-Einstein metric associated to 
a Sasaki-Einstein five-manifold. The corresponding Sasaki-Einstein five-metric is
\bea\label{SE5}
g_{\mathrm{SE}_5} &=& (\diff\varphi+A_{\mathrm{KE}})^2 + g_{\mathrm{KE}}~,
\eea 
which leads to a Calabi-Yau four-fold product metric on $\mathrm{CY}_3\times \C$ given by
\bea
g_{\mathrm{CY}_4} &=& \diff \rho_1^2 + \rho_1^2\left[ (\diff\varphi+A_{\mathrm{KE}})^2 + g_{\mathrm{KE}}\right] + 
\diff\rho_0^2+ \rho_0^2\diff\varphi_0^2~.
\eea
Here $\rho_0,\rho_1\in [0,\infty)$ are radial variables, and $\varphi_0$ has period $2\pi$. The corresponding Sasaki-Einstein seven-metric 
at unit distance from the conical singularity at $\{\rho_0=\rho_1=0\}$ is
\bea\label{SE7}
g_{\mathrm{SE}_7} &=& \frac{1}{1-r^2}\diff r^2 + r^2\left[(\diff\varphi+{A}_\mathrm{KE})^2 + g_\mathrm{KE}\right] +
(1-r^2)\diff\varphi_0^2~,
\eea
where $0\leq r\leq 1$. Note that the Killing vector fields $\partial_\varphi$ and $\partial_{\varphi_0}$ vanish 
at $r=0$ and $r=1$, respectively, and that the Reeb vector field is the sum
$\xi=\partial_{\varphi}+\partial_{\varphi_0}$. The metric (\ref{SE7}) 
is singular at $r=0$ (which is an $S^1$ locus parametrized by $\varphi_0$) unless the original Sasaki-Einstein five-manifold is 
$S^5$ equipped with its standard round metric. This is simply because 
the Calabi-Yau four-fold is also singular along $r=0$, which is the conical singularity of 
$\mathrm{CY}_3$.

It is no coincidence that the Sasaki-Einstein metric (\ref{SE7})
resembles our general metric (\ref{7dmetric}). The AdS$_4\times \mathrm{SE}_7$ 
background is the infrared limit of $N$ M2-branes at the conical singularity $\{\rho_0=\rho_1=0\}$ of $\mathrm{CY}_3\times \C$. 
The holomorphic function $z_0=\rho_0\ex^{\ii\varphi_0}$ leads to a scalar Kaluza-Klein mode 
on the Sasaki-Einstein seven-space, which in turn is dual to a gauge-invariant scalar chiral primary operator $\mathcal{O}$
in the dual three-dimensional SCFT. We may then consider deforming the SCFT by adding the 
operator $\lambda \mathcal{O}^p$. In three dimensions, this is a relevant deformation for $p=2$ and $p=3$, as discussed in 
 \cite{Jafferis:2011zi}.  Moreover, such a term can appear in the superpotential of a putative infrared fixed point also only if 
 $p=2$, $p=3$, since otherwise one violates the unitarity bound -- the R-charge/scaling dimension 
 of $\mathcal{O}$ would be $\Delta(\mathcal{O})=2/p$, and necessarily we have $\Delta(\mathcal{O})\geq \tfrac{1}{2}$ for a 
 unitary CFT in three dimensions, with equality only for a free field. The gravity dual to the infrared fixed point of the 
 massive $p=2$ deformation is the Corrado-Pilch-Warner solution of the previous section, 
 while we will find the $p=3$ solution as a numerical solution to the ODEs (\ref{ODEs}) in the next section.

In  \cite{Jafferis:2011zi} the authors studied $d=3$, $\mathcal{N}=2$ supersymmetric field theories for $N$ M2-branes 
on $\mathrm{CY}_3\times \C$ backgrounds, in particular computing 
the free energy using localization and matrix model techniques. 
This allows one to compute the ratio of UV and IR free energies, 
where the UV theory is dual to the AdS$_4\times \mathrm{SE}_7$ 
background, while the IR theory is the fixed point of the renormalization group flow 
induced by the $\lambda\mathcal{O}^p$ deformation. They found the universal formula, independent 
of the choice of CY$_3$,
\bea\label{fieldp}
\frac{\mathcal{F}_{\mathrm{IR}}}{\mathcal{F}_{\mathrm{UV}}} &=& \frac{16(p-1)^{3/2}}{3\sqrt{3}p^2}~.
\eea

We now show that this field theory result is easily obtained using our contact volume formula (\ref{freeenergy}), 
thus acting as a check of the AdS/CFT duality for this class of theories.
The CY$_3\times \C$ Calabi-Yau four-fold has at least a $\C^*\times\C^*$ symmetry,
in which the first $\C^*$ acts on the CY$_3$, and under which the CY$_3$ Killing spinors have charge $\tfrac{1}{2}$, 
and the second $\C^*$ acts in the obvious way on the copy of $\C$ with coordinate $z_0$. Let us denote the 
components of the Reeb vector field in this basis as $(\xi_1,\xi_0)$. In terms of the explicit 
coordinates introduced above, this gives the Reeb vector field as
\bea\label{Reebbasis}
\xi &=& \frac{1}{3} \xi_1\partial_{\varphi} + \xi_0\partial_{\varphi_0}~.
\eea
For the Calabi-Yau four-fold metric, we have already noted that $\xi_1=3$ and $\xi_0=1$.
In general, the Killing spinors have charge $2$, as in equation (\ref{charge2}), 
precisely when
\bea\label{weight4}
\xi_1+\xi_0 &=& 4~,
\eea
which is also equivalent to the holomorphic $(4,0)$-form $\Omega_{(4,0)}=\Omega_{(3,0)}\wedge \diff z_0$ having charge 4. 
As shown in appendix B of \cite{Gabella:2010cy}, in general the contact volume is a function of the Reeb vector field.
In our case the contact volume of $Y_7$ is given by the general formula
\bea\label{volumes}
\mathrm{Vol}(Y_7)[\xi_1,\xi_0] &=& \frac{1}{\xi_0}\, \mathrm{Vol}(Y_5)[\xi_1]~,
\eea
where $Y_5$ denotes the five-manifold link of CY$_3$. 
Using
$\xi_1=3$ for a Sasaki-Einstein metric,  (\ref{volumes}) implies the relation $\mathrm{Vol}(\mathrm{SE}_7)= \mathrm{Vol}(\mathrm{SE}_5)$ between Sasaki-Einstein volumes. Notice that $\xi_0=1$ 
 gives the expected scaling dimension $\Delta(\mathcal{O})=\tfrac{1}{2}$ of a free chiral field.\footnote{There is a factor of $\tfrac{1}{2}$ in going from the geometric scaling dimension under the Euler vector to the scaling dimension $\Delta$ in field theory, 
{\it cf.} equation (2.31) of \cite{Gauntlett:2006vf}.}

Let us now consider the IR solution corresponding to the deformation by $\lambda\mathcal{O}^p$. The 
scaling dimension of $\mathcal{O}$ necessarily changes from $\Delta(\mathcal{O})=\tfrac{1}{2}$ to $\Delta(\mathcal{O})=2/p$. 
Since the coordinate $z_0$ gives rise to the Kaluza-Klein mode leading to this BPS operator, this means 
the charge of $z_0$ under the Reeb vector field at the IR fixed point should be $\xi_0=4/p$. From (\ref{weight4}) we thus have $\xi_1=4(p-1)/p$. We then compute the contact volumes
\bea
\mathrm{Vol}(Y_7^{(p)}) &=& \frac{1}{\xi_0}\mathrm{Vol}(Y_5)[\xi_1] \ = \ \frac{1}{\xi_0}\left( \frac{\xi_1}{3}\right)^{-3}\mathrm{Vol}(Y_5)[3]\nonumber\\
&=& \frac{27p^4}{256(p-1)^3}\mathrm{Vol}(\mathrm{SE}_7)~.
\eea
Here we have used that the volume of a contact five-manifold 
is homogeneous degree $-3$ in the Reeb vector field \cite{Martelli:2006yb}, \cite{Gabella:2010cy}.
Taking the square root and using our free energy formula (\ref{freeenergy}), we precisely reproduce the field theory result (\ref{fieldp})!\footnote{
Notice for $p\geq 4$ this is a somewhat formal agreement, since 
the IR fixed point is not expected to exist due to the unitarity bound, 
as explained above.}

We conclude by recording that the Reeb vector field (\ref{Reebbasis}) at the IR fixed point  is
\bea\label{Reebp}
\xi &=& \frac{4(p-1)}{3p}\partial_{\varphi} + \frac{4}{p}\partial_{\varphi_0}~.
\eea
This will be crucial in the following sections when we consider the appropriate boundary conditions
for the ODEs (\ref{ODEs}).

\subsection{The Corrado-Pilch-Warner solution (again)}

Before moving on to  the gravity dual of the cubic $p=3$ deformation, let us consider again the 
explicit $p=2$ Corrado-Pilch-Warner solution. The analysis in the previous section implies that the Reeb vector field 
should be
\bea\label{CPWReeb}
\xi &=& 4\partial_\psi \ = \ \frac{2}{3}\partial_{\varphi} + 2\partial_{\varphi_0}~,
\eea
where $\psi$ is the coordinate in (\ref{7dmetric}). This fact is very closely related 
to the appropriate boundary conditions one needs to impose on the ODEs (\ref{ODEs}) 
in order to obtain a good supergravity solution. For the explicit solution in section \ref{sec:CPW}, the coordinate $r\in[0,2\sqrt{2}]$, 
and by definition $\partial_{\varphi_0}$ is the Killing vector field 
that vanishes at $r=0$, while $\partial_{\varphi}$ vanishes at $r=2\sqrt{2}$. 
Let us see how this works precisely.  
Without loss of generality we henceforth set 
\bea
\gamma & = & 3~.
\eea

Near to $r=0$, we may use $f(0)=2\gamma$, $\alpha(r)=2^{-1/4}r^{-1/2}+\mathcal{O}(r^{1/2})$ to compute
\bea
\|A\partial_\psi + B\partial_\tau\|^2\mid_{r=0} & =  & \frac{1}{16}\left(A+2\gamma B\right)^2~.
\eea
This vanishes only if $A=-2\gamma B$, so that the vanishing vector field  at $r=0$ is
\bea
\partial_{\varphi_0} &\propto & -2\gamma \partial_\psi + \partial_\tau~.
\eea
To determine the proportionality constant we need to examine the rate of collapse. Introducing $r=4\sqrt{2}R^2$, 
we have near to $r=0$ that
$\frac{\alpha^2}{16}\diff r^2=   \diff R^2[1+\mathcal{O}(R^2)]$. Thus $R$ measures 
geodesic distance from $R=0$, to leading order, and if $\partial_{\varphi_0}$ 
is such that $\varphi_0$ has period $2\pi$ and $\partial_{\varphi_0}$ vanishes at $R=0$, then 
the metric will be smooth here only if $\|\partial_{\varphi_0}\|=R$. 
Said another way, to leading order near to $R=0$ the metric must be the 
standard metric $\diff R^2 + R^2 \diff\varphi_0^2$ on $\R^2$ in polar coordinates $(R,\varphi_0)$.
We then compute
\bea
\|-2\gamma\partial_\psi + \partial_\tau\|^2 & = & {\gamma^2 R^2} +\mathcal{O}(R^4)~.
\eea
This fixes
\bea\label{CPWphi0}
\partial_{\varphi_0} &=& 2\partial_\psi - \frac{1}{\gamma}\partial_\tau~.
\eea

We may perform a similar analysis near to $r=2\sqrt{2}$.
Introducing $2\sqrt{2}-r \equiv 4\sqrt{2}Z^2$, we have
%$f= \gamma s/\sqrt{2}$, 
$f= 4\gamma Z^2$, while near to $Z=0$ we have $\alpha= 2^{-3/2}Z^{-1}+\mathcal{O}(Z)$. 
Now
\bea
\|A\partial_\psi + B\partial_\tau\|^2 &= & \frac{1}{16\cdot 9}\left[9A^2+\mathcal{O}(\nes^2)\right]~,
\eea
so this vector field vanishes at $r=2\sqrt{2}$ only if $A=0$, leading to 
\bea
\partial_{\varphi} &\propto & \partial_\tau~.
\eea
%\textcolor{red}{$S$ CHANGED TO $\nes$}
 %Again, to compute the coefficient we introduce $s=4\sqrt{2}\nes^2$ and 
In particular, the coefficient may be computed from
$\frac{\alpha^2}{16}\diff r^2  = \diff \nes^2[1+\mathcal{O}(\nes^2)]$ and 
\bea
\|\partial_\tau\|^2 & = & \frac{\gamma^2}{9}\nes^2 \ = \ \nes^2~,
\eea
where we have  used $\gamma=3$ in the last step. This is indeed the expected result, since for the canonical scaling of 
$\gamma=3$ the connection term $\diff\tau + \mathbb{A}_\mathrm{KE}$ in the 
metric (\ref{7dmetric})  must be the contact one-form $\diff\varphi + \mathbb{A}_\mathrm{KE}$ for the original Sasaki-Einstein five-manifold 
(\ref{SE5}), implying that indeed $\partial_{\tau} = \partial_{\varphi}$. The collapsing part of the metric near to $r=2\sqrt{2}$ is then 
$\diff \nes^2 + \nes^2((\diff\tau+\mathbb{A}_\mathrm{KE})^2 + g_{\mathrm{KE}})$.
This locally is precisely the CY$_3$ conical metric, giving a smooth collapse 
at $\nes=0$ if and only if the the K\"ahler-Einstein metric is the standard metric on $\mathbb{CP}^2$. 
More generally, $r=2\sqrt{2}$ is an $S^1$ locus of CY$_3$ cone singularities.

To summarize, putting (\ref{CPWphi0}) together with $\partial_{\tau} = \partial_{\varphi}$ we have shown 
\bea
2\partial_\psi &=&  \partial_{\varphi_0} + \frac{1}{3}\partial_{\varphi}~.
\eea
Recalling that the Reeb vector field is $\xi=4\partial_\psi$, we have thus shown
\bea
\xi &=& 4\partial_\psi \ = \ \frac{2}{3}\partial_{\varphi}+2\partial_{\varphi_0}~.
\eea
This precisely coincides with (\ref{CPWReeb}), which was derived in the previous section 
based only on topological and scaling arguments.

\subsection{Cubic deformations}

\label{newsolution}

We may now use precisely the same arguments as the previous section to deduce the 
appropriate boundary conditions for the ODEs (\ref{ODEs}) in the case of cubic $p=3$ deformations. 
The Reeb vector field is now
\bea
\xi &=& 4\partial_\psi \ = \ \frac{8}{9}\partial_{\varphi}+ \frac{4}{3}\partial_{\phi_0}~,
\eea
where  by definition again $\partial_{\phi_0}$ and $\partial_{\varphi}$ are the vanishing 
vector fields, while $\psi$ is the coordinate in our metric (\ref{7dmetric}).

Let us begin by considering the behaviour near to $r=0$. Suppose that $\alpha(r)=wr^{\nu}+o(r^{\nu})$, with 
$w$ a non-zero constant. Then the first ODE in (\ref{ODEs}) implies
\bea
(\log f)' & \sim & -\frac{w^2}{2} r^{1+2\nu}~,
\eea
which leads to the leading order solution 
\bea
f(r)\sim A_0 \exp\left[-\frac{w^2r^{2(1+\nu)})}{4(1+\nu)}\right]~,\eea
where $A_0$ is a constant.
The second ODE in (\ref{ODEs})
is then to leading order
\bea
\gamma &\sim & \frac{A_0 w r^\nu (-\nu + w^2r^{2(1+\nu)}) \exp\left[-\frac{w^2r^{2(1+\nu)})}{4(1+\nu)}\right]}{\sqrt{1+w^2r^{2\nu}(1+r^2)}}~.
\eea
For $\nu>0$ the right hand side tends to zero as $r\rightarrow 0$, which is a contradiction. This is also the case 
for $\nu=0$. On the other hand, $f(r)$ blows up exponentially at $r=0$ unless $\nu>-1$. 
Since we do not want the size of the K\"ahler-Einstein 
metric to blow up on $Y_7$, a regular solution must hence have
$-1<\nu<0$. Given this, to leading order the last equation becomes
\bea
\gamma &\sim & -A_0\nu w \left(r^{-2\nu}+w^2\right)^{-1/2} \ \stackrel{r\rightarrow 0}{\longrightarrow} \ -A_0\nu~.
\eea
Thus we conclude that $3=\gamma=-A_0\nu$. 
Note that $A_0>0$, and that the metric (\ref{7dmetric}) is then positive definite only if $w>0$.

As in the previous section, introducing $r=\left(\frac{4(1+\nu)}{w}\right)^{1/(1+\nu)}R^{1/(1+\nu)}$
we compute
\bea
\frac{\alpha^2}{16}\diff r^2 &\sim & \frac{w^2 r^{2\nu}\diff r^2}{16} \ = \ \diff R^2~,
\eea
We now determine the vanishing vector field at $r=0$, computing 
\bea\label{bob}
\|A\partial_\psi + B\partial_\tau\|^2\mid_{R=0} & = & \frac{1}{16}\left(A-\frac{B\gamma}{\nu}\right)^2~,
\eea
where we have eliminated $A_0=-\gamma/\nu$. Thus the vector field
$-\frac{1}{\nu}\partial_\psi-\frac{1}{\gamma}\partial_\tau$
vanishes at $r=0$. To fix the normalization we need the rate of collapse:
\bea
\left\|-\frac{1}{\nu}\partial_\psi -\frac{1}{\gamma}\partial_\tau\right\|^2&=& \frac{(1+\nu)^2}{\nu^2}R^2+o(R^2)~,
\eea
near to $R=0$. This fixes
\bea
\partial_{\varphi_0} &=& \frac{1}{1+\nu}\partial_\psi  +\frac{\nu}{\gamma(1+\nu)}\partial_\tau~.
\eea
In fact this is already enough to determine $\nu$. Recall that $\xi=4\partial_\psi$ is the Reeb vector field, so we can also write
\bea
\partial_{\varphi_0} &=& \frac{1}{4(1+\nu)}\xi  +\frac{\nu}{\gamma(1+\nu)}\partial_\tau~.
\eea
Since the coordinate $z_0$ on $\C$ has charge $2/p$ under $\xi$, we thus conclude that in general
\bea
1 &=& \frac{1}{4(1+\nu)} \cdot \frac{4}{p}~,
\eea
so that
\bea
\nu &=& -1 + \frac{1}{p}~.
\eea
In particular, the Corrado-Pilch-Warner solution has $\nu=-\frac{1}{2}$, while for the 
cubic deformation we should set $\nu = -\frac{2}{3}$. The boundary condition for $\alpha(r)$ near to $r=0$ is in general $\alpha(r)\sim w r^{-1+1/p}$. 
 It is important to note that, with this boundary condition on $\alpha(r)$, the metric is  
 completely smooth near to $r=0$. Although $\alpha(r)$ is blowing up, the function 
 $\alpha f/\sqrt{1+\alpha^2(1+r^2)}\sim f(0) = -\gamma/\nu$, so that the K\"ahler-Einstein 
 factor in (\ref{7dmetric}) has finite non-zero size. The remaining Killing vector that is not zero also has finite length 
 at $r=0$, as one sees from (\ref{bob}). 
 
 We can now similarly analyse the other collapse. This is necessarily at a zero of $f(r)$. 
To see this, note that the K\"ahler-Einstein part of the metric (\ref{7dmetric}) collapses at 
either a zero of $\alpha$, or a zero of $f$ (potentially both). Suppose this is at $r=r_0$. If $\alpha\sim
\upsilon(r_0-r)^\eta$ to leading order, with $\eta>0$, then solving 
the ODE for $f$ leads to the leading order result
\bea
f(r) &\sim & A_1\exp\left[\frac{\upsilon^2r_0(r_0-r)^{1+2\eta}}{2(1+2\eta)}\right]~.
\eea 
Thus $f(r_0)=A_1$ is in fact non-zero. The second ODE in (\ref{ODEs}) is then 
consistent near to $r=r_0$ only if the exponent $\eta=1$, which means 
that $\alpha(r)\sim \upsilon (r_0-r)$ is a simple zero. However, from the metric (\ref{7dmetric}) 
we see that in fact then the entire metric collapses at $r=r_0$, which does not give the 
correct topology. So we can rule out $\alpha(r)$ having a zero at $r=r_0$.

Thus $f(r_0)=0$. Let us suppose that to leading order
\bea
f(r)& \sim & q (r_0-r)^{\lambda}~,\eea
with $\lambda>0$. Then from the first ODE in (\ref{ODEs}) we obtain
\bea
\alpha(r)& \sim & \sqrt{\frac{2\lambda}{r_0(r_0-r)}}~.
\eea
Notice that for the Corrado-Pilch-Warner solution we have
$\lambda_{\mathrm{CPW}}=1$, and this leading order solution for $\alpha(r)$ near to 
$r=r_0$ is in fact the \emph{exact} solution. 
For our cubic $p=3$ solution $\alpha(r)$ must instead interpolate between $r^{-2/3}$ behaviour near to $r=0$ 
and $(r_0-r)^{-1/2}$ behaviour near to $r=r_0$.
The second ODE again fixes the exponent $\lambda=1$ for consistency near to $r=r_0$, and we conclude that
\bea
f(r)& \sim & q (r_0-r)~,\\
\alpha(r)& \sim & \sqrt{\frac{2}{r_0(r_0-r)}}~,
\eea
near to $r=r_0$. Moreover, the second ODE then fixes
\bea\label{gammaq}
\gamma &=& \frac{3qr_0}{2\sqrt{1+r_0^2}}~.
\eea

Finally, we turn to looking at the vanishing vector field. 
Writing $r_0-r \equiv  2r_0\wes^2$, 
we find that $\frac{\alpha^2\diff r^2}{16}\sim \diff \wes^2$. Then
\bea\label{bobagain}
\|A\partial_\psi + B\partial_\tau\|^2 & = & \frac{1}{16}A^2 + \mathcal{O}(\wes^2)~,
\eea
so that the vanishing vector field at the root $r=r_0$ is again proportional to 
$\partial_\tau$. We find more precisely that, quite remarkably,
\bea
\|\partial_{\tau}\|^2 & = & \left(\frac{\gamma}{3}\right)^2\wes^2+o(\wes^2)~,
\eea
where we have substituted for $q$ using (\ref{gammaq}). This is exactly the same 
behaviour as for the Corrado-Pilch-Warner solution near to this root.
Since this collapsing vector field is by definition $\partial_{\varphi}$, we again conclude that
\bea
\partial_{\tau} &=& \partial_{\varphi}~.
\eea
Again, this had to be the case for global reasons associated to the 
form of the connection one-form appearing in the metric. Again one finds that 
$r=r_0$ is an $S^1$ family of CY$_3$ cone singularities, with the analysis being 
identical to that for the Corrado-Pilch-Warner solution in the previous section.

This completes our analysis of the regularity conditions. Setting $\gamma=3$, we have shown that the Reeb vector field is
\bea
\xi &=& -\frac{4\nu}{3}\partial_{\varphi}+4(1+\nu)\partial_{\phi_0}~.
\eea
Using the fact that $\nu=-1+\frac{1}{p}$, this precisely agrees with our topological analysis in section \ref{sec:def}, 
and in particular the formula (\ref{Reebp}).

\subsection{Summary and numerics}

We may summarize the results of the previous sections as follows. 

\begin{quote}The gravity dual to the infrared fixed point of a deformation 
of a CY$_3\times \C$ background by the operator $\lambda \mathcal{O}^p$ may be obtained by solving the coupled 
set of ODEs for $\alpha(r)$, $f(r)$:
\bea
 f' &=& -\frac{r\alpha^2}{2}f~,\nonumber\\
 \frac{(r\alpha'-r^2\alpha^3)f}{\sqrt{1+(1+r^2)\alpha^2}} &=& -3~.
 \eea
 The boundary conditions are that near to $r=0$ we have $\alpha(r)\sim wr^{-1+1/p}$, with $w>0$ 
 a constant. Using the second ODE above this implies that  $f(0)=3p/(p-1)$. 
 Then near to $r=r_0$, for some $r_0>0$, we must impose that $f(r)\sim q(r_0-r)$, where 
 the ODEs imply that $\alpha(r)\sim \sqrt{2/r_0(r_0-r)}$ and  $q=2\sqrt{1+r_0^2}/r_0$. 
 With these boundary conditions we obtain a smooth supergravity solution, up to the 
expected $S^1$ locus of CY$_3$ singularities along $r=r_0$. When the CY$_3$ is 
simply flat $\C^3$, in particular we obtain a completely smooth $\mathcal{N}=2$ 
supergravity solution with the topology AdS$_4\times S^7$.
\end{quote}

The Corrado-Pilch-Warner solution precisely  solves this problem for $p=2$, and physical arguments imply 
there should also be a solution for $p=3$. We have not been able to find this solution analytically, but 
it is straightforward to solve the ODEs numerically with the above boundary conditions. 

We first change variable to $r=R^3$, and then solve the second order ODE (\ref{2ndorder}) in a Taylor 
expansion in $R$, around $R=0$, up to some large order. Using the constraint that $f(0)=3p/(p-1)=9/2$ we find
\bea\label{Taylor}
f(R) \ =\  \frac{9}{2} - cR^2 - \frac{c^2}{9}R^4 + \frac{2187-128c^3}{3888}R^6 + \frac{19683 c + 1264 c^3}{104976}R^8 + \mathcal{O}(R^{10})~,
\eea
where $c$ is an arbitrary integration constant. This then implies
\bea
\alpha(R) &=&\frac{2}{3}\sqrt{\frac{2}{3}} c^{1/2}R^{-2} + \frac{4}{27}\sqrt{\frac{2}{3}}c^{3/2} - 
\frac{(2187-224 c^3)}{1944\sqrt{6}} c^{-1/2}R^2 + \mathcal{O}(R^4)~.
\eea
Thus $\alpha(r)$ has the correct behaviour   $\alpha(r)\sim wr^{-2/3}$, where we identify 
the constant $w= \sqrt{8c/27}$.

We then have a numerical shooting problem: for each choice of integration constant $c$, we 
solve the second order ODE (\ref{2ndorder}) (or equivalently the coupled first order system), 
with initial Taylor expansion (\ref{Taylor}). 
We simply require that $f(r_0)=0$ for some $r_0>0$. From the analysis in the previous section, 
the ODEs themselves imply that a zero of $f(r)$ is automatically a simple zero. 
\begin{figure}[ht!]
\centering
\epsfig{file=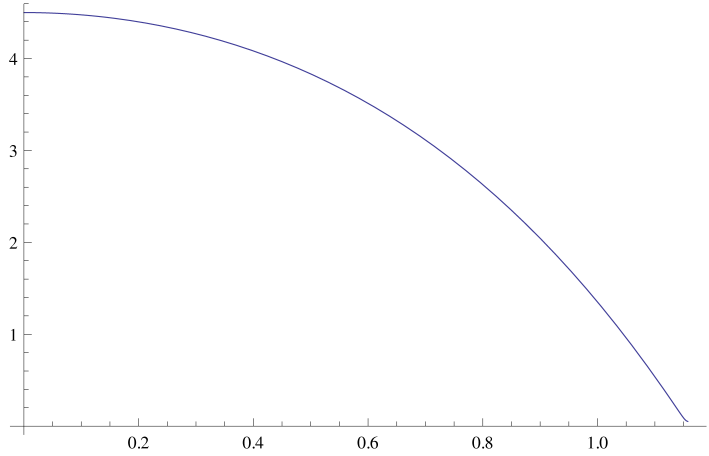,width=0.6\textwidth}
\caption{Numerical plot of the function $f(R)$ with integration constant $c\simeq 2.4998$. 
Note that $f(0)=9/2$ and $f(R)$ decreases monotonically to zero at $R=R_0$, where $R_0\simeq 1.16$. }\label{fig1}
\end{figure}

\begin{figure}[ht!]
\centering
\epsfig{file=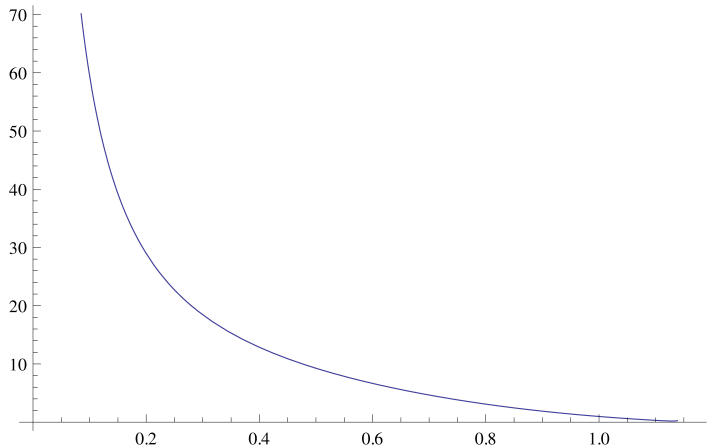,width=0.6\textwidth}
\caption{Numerical plot (with integration constant $c\simeq 2.4998$) of the function $f'(R)+\frac{6\sqrt{1+R^6}}{R}$, 
which should tend to zero at $R=R_0\simeq 1.16$.}\label{fig2}
\end{figure}

We find that there exists a point $r_0>0$ with $f(r_0)=0$ for the choice
\bea
c& \simeq & 2.4998~.
\eea 
The resulting plot of the function $f(R)$, with $R=r^3$, is shown in Figure \ref{fig1}. 
Smaller values of $c$ lead to $f(R)$ remaining positive, while for $c>2.4998$ we 
find the numerics becomes highly unstable. Indeed, 
the numerics is slightly unstable near the zero of $f$ for $c=2.4998$. As a cross check that we really do have 
a zero,  we note that at a zero of $f(R)$ we necessarily have
\bea
f'(R_0) &=& -\frac{6\sqrt{1+R_0^6}}{R_0}~.
\eea
In Figure \ref{fig2} we numerically plot the function $f'(R)+\frac{6\sqrt{1+R^6}}{R}$, which 
should tend to zero at $R=R_0$.

Of course, it is quite tantalizing that the numerical value of $c$ is so close to $5/2$, 
perhaps suggesting the possibility of an analytic solution, or at least an analytic 
explanation of $c=5/2$. We leave this question open.

\section{Conclusions}

\label{concsec}

The main result of this paper is the determination of the necessary and sufficient conditions
 on supersymmetric solutions of $D=11$ supergravity that are dual to ${\cal N}=2$
three-dimensional superconformal field theories. The eleven-dimensional metric is taken
to be a warped product of AdS$_4$ with a seven-dimensional Riemannian metric, and we have
allowed for the most general four-form $G$ consistent with $SO(3, 2)$ symmetry. We showed
that generically the supersymmetry conditions may be formulated in terms of a canonical 
local $SU(2)$-structure on the seven-dimensional manifold $Y_7$. The well-known 
Freund-Rubin AdS$_4\times Y_7$ solutions where $Y_7$ is Sasaki-Einstein  arise as a special case, characterized
by an $SU(3)$-structure. For solutions with non-zero M2-brane charge, we showed that 
many geometrical and physical properties of $Y_7$ are captured by a contact structure, elaborating on the
results presented in \cite{Gabella:2011sg}. We also recovered the class of general solutions
with vanishing M2-brane charge, previously discussed in \cite{Gauntlett:2006ux}.
 
By imposing a single additional requirement, that a certain vector bilinear is a Killing vector, 
we reduced the conditions to solving a second order non-linear ODE. The seven-dimensional metric
on $Y_7$ is then fully specified by the choice of a (local) four-dimensional K\"ahler-Einstein 
metric, and any solution to this ODE. We managed to find an analytic solution of the 
ODE, and showed that this reproduces a class of solutions found originally in \cite{Corrado:2001nv}. 
In addition, using a combination of analytic and numerical methods, we have discovered a 
further solution to our ODE, 
yielding a class of new supersymmetric AdS$_4$ solutions with non-trivial four-form flux. 
These can be interpreted as holographically dual to certain cubic superpotential 
deformations of ${\cal N}=2$ Chern-Simons gauge theories. When the K\"ahler-Einstein metric
is chosen to be that on $\mathbb{CP}^2$, the seven-dimensional metric is a smooth (non-Einstein) 
metric on $S^7$, different from that of \cite{Corrado:2001nv}. We suspect that there are no further 
regular solutions in this class.
%, and it would be nice to prove this.

Our work may be regarded as providing the foundation for studying more general aspects of 
${\cal N}=2$ three-dimensional superconformal field theories with M-theory duals. For example,
we expect that the geometric characterization of solutions we presented may be used to
attack general problems, such as the gravity dual of ${\cal F}$-maximization, similarly to 
the developments in \cite{Gabella:2009wu, Gabella:2010cy}. It is also clear that using our results
it will be possible to construct a consistent Kaluza-Klein truncation to four dimensions, 
extending that in \cite{Gauntlett:2007ma}.
The AdS$_4$ solutions dual to beta-deformations \cite{Lunin:2005jy, Gauntlett:2005jb} of ${\cal N}=2$ field theories must solve
the equations that we presented, and it would interesting to verify this explicitly.
Of course, it would also be very interesting to use our general equations  as a method 
for finding new solutions (perhaps numerically), outside the classes that have been discovered so 
far\footnote{These include, for example, the gravity duals of general ${\cal N}=2$ marginal deformations \cite{Kol:2002zt}.}.
These are all exciting directions for future work. 

\subsection*{Acknowledgments}
\noindent
We would like to thank the Isaac Newton Institute for hospitality.
M.~G. is supported by the ERC Starting Independent Researcher Grant
259133-ObservableString, D.~M. is supported by the EPSRC Advanced Fellowship EP/D07150X/3, A.~P. is supported by 
an A.~G. Leventis Foundation grant and via
the Act ``Scholarship Programme of S.~S.~F. by the procedure of individual
assessment, of 2011-12'' by resources of the Operational Programme for Education
and Lifelong Learning, of the European Social Fund (ESF) and of the NSRF,
2007-2013. D.~M. also acknowledges partial support from the STFC grant ST/J002798/1.
J.~F.~S. is supported by the Royal Society and Oriel College. 

\appendix

\section{Some useful identities}\label{app:useful}

In this appendix we collect a number of useful identities that have been used repeatedly to
derive the results presented in the main text.

From the algebraic equation in (\ref{spinoreqns}) one can derive the following useful identities
\bea
\left( \bar \chi_i^c \mathcal{C} \chi_j^c  + \bar \chi_i \mathcal{C} \chi_j  \right) - \frac{\ii m}{3}\ex^{-3\Delta} \bar \chi_i^c \mathcal{C} \chi_j~~~~~~~~~~~~~~~~~~~~~~~~~~~~~~~~~~~~~~~~~\nn\\ 
 + \frac{1}{2} \de_m \Delta \bar \chi_i^c [\mathcal{C}, \gamma^m]_-  \chi_j + \frac{1}{288} F_{mnpq}\ex^{-3\Delta} \bar \chi_i^c [\mathcal{C}, \gamma^{mnpq}]_+  \chi_j & = & 0~,~~~~
\label{useful1}\\
\left( \bar \chi_i^c \mathcal{C} \chi_j^c  - \bar \chi_i \mathcal{C} \chi_j  \right)  
+ \frac{1}{2} \de_m \Delta \bar \chi_i^c [\mathcal{C}, \gamma^m]_+  \chi_j + \frac{1}{288} F_{mnpq}\ex^{-3\Delta} \bar \chi_i^c [\mathcal{C}, \gamma^{mnpq}]_-  \chi_j & = & 0~,
\label{useful2}
\eea
where $\mathcal{C}\in $ Cliff$(7)$ is an arbitrary element of the Clifford algebra and $[~,~]_\pm$ denotes the (anti)-commutator.  Similarly
we note
\bea
\left( \bar \chi_i^c \mathcal{C} \chi_j  -  \bar \chi_i \mathcal{C} \chi_j^c  \right) + \frac{\ii m}{3}\ex^{-3\Delta} \bar \chi_i \mathcal{C} \chi_j~~~~~~~~~~~~~~~~~~~~~~~~~~~~~~~~~~~~~~~~~\nn\\ 
- \frac{1}{2} \de_m \Delta \bar \chi_i [\mathcal{C}, \gamma^m]_-  \chi_j - \frac{1}{288} F_{mnpq}\ex^{-3\Delta} \bar \chi_i [\mathcal{C}, \gamma^{mnpq}]_-  \chi_j & = & 0~,~~~~
\label{useful3}\\
 \left( \bar \chi_i^c \mathcal{C} \chi_j  + \bar \chi_i \mathcal{C} \chi_j^c  \right)  
 + \frac{1}{2} \de_m \Delta \bar \chi_i [\mathcal{C}, \gamma^m]_+  \chi_j + \frac{1}{288} F_{mnpq}\ex^{-3\Delta} \bar \chi_i [\mathcal{C}, \gamma^{mnpq}]_+  \chi_j & = & 0~.
\label{useful4}
\eea
Similar identities exist in the alternative basis (\ref{chipm}). 

From the Fierz identity for the $\mathrm{Cliff(7)}$ algebra 
\bea\label{fierz7}
\bar{\xi}_1 \xi_2 \, \bar{\xi}_3 \xi_4  & = &
\frac{1}{8} \Bigg[ \bar{\xi}_1 \xi_4 \, \bar{\xi}_3 \xi_2 
+ \bar{\xi}_1 \g_m \xi_4 \, \bar{\xi}_3 \g^m \xi_2 
- \frac{1}{2!} \bar{\xi}_1 \g_{mn} \xi_4 \, \bar{\xi}_3 \g^{mn} \xi_2 \nonumber\\
&&
- \frac{1}{3!} \bar{\xi}_1 \g_{mnp}\xi_4 \, \bar{\xi}_3 \g^{mnp} \xi_2 \Bigg]~,
\eea
where $\xi_a$, $a=1,2,3,4$, are arbitrary Spin(7) spinors,
we derive the useful identity
\bea\label{Fierz}
\bar{\xi^c_1} \g^m \xi_2 \, \bar{\xi^c_2} \g_m \xi_4 
 &=&  \bar{\xi^c_1} \xi_4 \, \bar{\xi^c_2} \xi_2 - \bar{\xi^c_1} \xi_2 \, \bar{\xi^c_2}  \xi_4~.
\eea

\section{$SU(2)$- and $SU(3)$-structures in dimension $d=7$}\label{app:structures}

In the main text we have presented our results, summarized in the equations in section \ref{sec:NS}, 
in terms of an $SU(2)$-structure. This is defined by the three one-forms $E_1$, $E_2$, $E_3$, and 
three $SU(2)$-invariant two-forms $J_1$, $J_2$, $J_3$. In arguing that the conditions 
we write are sufficient, it is also convenient to think of this in terms two $SU(3)$-structures,
defined by the Killing spinors $\chi_\pm$. In this appendix we present explicit 
formulas for the spinor bilinears in terms of both $SU(2)$- and $SU(3)$-structures.

\subsection{$SU(2)$-structure}

Recall that the $SU(2)$-structure is specified by two spinors $\chi_1$, $\chi_2$, or equivalently the 
linear combinations $\chi_\pm \equiv  \frac{1}{\sqrt{2}}\left(\chi_1\pm \ii \chi_2\right)$ defined 
in (\ref{chipm}). Here we choose to use $\chi_\pm$ as our basis.

We then have the following zero-form bilinears
\bea\label{0form}
\bar\chi_+\chi_+ & = &   \bar\chi_-\chi_- \ =\  1~,\nn\\
 \bar\chi_+\chi_- & = & 0~,\nonumber\\
S & \equiv & \bar\chi_+^c\chi_+ \  = \ (\bar\chi_-^c\chi_-)^*~,\nn\\
\zeta &\equiv & \ii\bar\chi_+^c\chi_-  \ = \ \frac{ m}{6}\ex^{-3\Delta}~,
\eea
one-form bilinears
\bea\label{1form}
K & \equiv & \ii \bar\chi_+^c\gamma_{(1)}\chi_-  \ = \ \|\xi\| E_1~, \nn\\
 L & \equiv & \bar\chi_-\gamma_{(1)}\chi_+ \ = \ \frac{S}{|S|}\left( \ii \frac{|S|}{\|\xi\|}E_1+\sqrt{1-\|\xi\|^2}E_2- \ii \frac{\zeta\sqrt{1-\|\xi\|^2}}{\|\xi\|}E_3\right)~,\nonumber\\
P &\equiv & -\bar\chi_+\gamma_{(1)}\chi_+  \ = \  \bar\chi_-\gamma_{(1)}\chi_- \ = \ \frac{\zeta}{\|\xi\|}E_1 + \frac{|S|\sqrt{1-\|\xi\|^2}}{\|\xi\|}E_3~,
\eea
two-form bilinears
\bea
V_\pm & \equiv & \frac1{2\ii} \left[ \bar \chi_+\gamma_{(2)} \chi_+ \pm  \bar\chi_- \gamma_{(2)} \chi_- \right] ~,\nn\\
V_+ &=&   \sqrt{1-\|\xi\|^2} J_2~, \nn\\
V_- &=&  \zeta J_3 + \frac1{\|\xi\|} E_2 \wedge\left( |S| \sqrt{1-\|\xi\|^2}  E_1 - \zeta E_3 \right)~, \nn\\
\bar\chi^c_+ \gamma_{(2)} \chi_- &=& - J_3 + \|\xi\| E_2 \wedge E_3 - \ii \sqrt{1-\|\xi\|^2} J_1
\label{deftwobil}
\eea
and three-form bilinears
\bea
W_\pm & \equiv & \frac{1}{2}\left[\bar\chi_+^c\gamma_{(3)}\chi_+ \pm \left(\bar\chi_-^c\gamma_{(3)}\chi_-\right)^*\right]~,\nn\\
\Real\left[\frac{|S|}{S}W_-\right] &=& -  \sqrt{1-\|\xi\|^2} J_3 \wedge E_2, \nn \\
\Imag\left[\frac{|S|}{S}W_-\right] &=& -  J_3 \wedge \left(  \frac{|S| }{\|\xi\|}E_1-  \frac{\zeta \sqrt{1-\|\xi\|^2}}{\|\xi\|}E_3\right) + |S| E_{123}~, \nn \\
\Real\left[\frac{|S|}{S}W_+\right] &=& \frac{1}{\|\xi\|}J_1 \wedge \left( |S| \sqrt{1-\|\xi\|^2} E_1 - \zeta E_3\right) + \zeta J_2 \wedge E_2 ~, \nn\\
\Imag\left[\frac{|S|}{S}W_+\right] &=& - J_1 \wedge E_2 -\|\xi\| J_2 \wedge E_3 ~,\nn\\
\mathrm{Im}\left[ \bar\chi_+^c \gamma_{(3)}\chi_-\right] &= & |S| J_2 \wedge E_2 - \frac{1}{\|\xi\|}J_1 \wedge (\zeta\sqrt{1-\|\xi\|^2}E_1 + |S|E_3) ~,\nn\\
\ii \bar\chi_+ \gamma_{(3)} \chi_+ & =& \|\xi\| J_3 \wedge E_1 - E_{123} + |S| J_1 \wedge E_2 \nn\\
&& +\frac{1}{\|\xi\|} J_2 \wedge \left( \zeta \sqrt{1-\|\xi\|^2} E_1 + |S| E_3\right)~.
\eea
Notice that the two-forms and three-forms above are an incomplete list -- we have included only those bilinears that 
are referred to explicitly in the text.

\subsection{$SU(3)$-structures}

Recall that we defined the two non-canonical $SU(3)$-structures as real vectors $\mathcal{K}_\pm\equiv \bar\chi_\pm\gamma_{(1)}\chi_\pm$, real two-forms $\mathcal{J}_\pm\equiv -\ii\bar\chi_\pm\gamma_{(2)}\chi_\pm$, 
and complex three-forms $\Omega_\pm\equiv\bar\chi_\pm^c\gamma_{(3)}\chi_\pm$. Then we have the one-form bilinears
\bea
\mathcal{K}_+ &=& - \mathcal{K}_- \ = \  - P~,
\eea
two-form bilinears
\bea
\mathcal{J}_\pm &=& V_+\pm V_-~,
\eea
and three-form bilinears
\bea
\Omega_+ &=& W_++W_-~, \qquad \Omega_- \ = \ (W_+-W_-)^*~.
\eea

\section{The Sasaki-Einstein case}\label{app:SE}

In this appendix we study the case in which the three one-forms $K, \mathrm{Re}\, S^*L, \mathrm{Im}\, S^*L$ 
are linearly dependent. When they are linearly independent we have an $SU(2)$ structure, and in an open set we
can then introduce corresponding coordinates, as described in section \ref{sec:frame}. 
Since these one-forms are derived from spinor bilinears, linear dependence implies 
we have an $SU(3)$ structure. Focusing on the $m\neq 0$ case for clarity, 
we will prove that the only solutions for which we have a global $SU(3)$ 
structure are Sasaki-Einstein.

In order to proceed, we impose the linear relation
\bea\label{depend}
a K + b\, \Real S^*L + c\, \Imag S^* L &=& 0~,
\eea
with $a$, $b$, $c$ not all zero. Making use of the Fierz identity in (\ref{Fierz}) it is 
straightforward to compute the dot products of each of $K, \mathrm{Re}\, S^*L, \mathrm{Im}\, S^*L$ 
into this equation. An analysis of the resulting three equations then implies 
that at least one of $|S|=0$ or $\|\xi\|=1$ must hold. In particular, if $|S|=0$ 
then necessarily $a=0$, while if $\|\xi\|=1$ then $a=c(\zeta^2-1)$. 
The following analysis then treats these cases in turn.

If $|S|=0$ then of course also $S=0$. The bilinear equation (\ref{Sbilinear})
then implies that $L=0$ and hence in particular that the one-form
$\bar{\chi}_1\gamma_{(1)}\chi_1=0$. This says that $\chi_1$ defines 
a $G_2$ structure, rather than an $SU(3)$ structure, and hence that 
$\chi_1$ satisfies a reality (Majorana) condition $\chi_1=\mu \chi_1^c$. 
The scalar bilinears determine that $\mu=-\ii/\zeta$, and since 
$|\mu|^2=1$ we conclude that $\zeta=1$ and the warp factor is 
constant $\ex^{3\Delta}=m/6$. Finally, the bilinear equation 
(\ref{*Feqn}) and its $\chi_-$ analogue  imply
\bea
\ex^{3\Delta}\star F &=& \diff \left( \ii\, \ex^{6\Delta}\bar{\chi}_1\gamma_{(2)}\chi_1\right)  - 6\ex^{6\Delta}\mathrm{Im}\left[ \bar\chi_1^c \gamma_{(3)}\chi_1\right]~,
\eea
which in turn immediately implies that $F=0$.  This is because the Majorana condition
$\chi_1=-\ii \chi_1^c$ implies that the two-form bilinear $\bar{\chi}_1\gamma_{(2)}\chi_1=0$ 
(there are no $G_2$-invariant two-forms), while the three-form bilinear 
$\bar\chi_1^c \gamma_{(3)}\chi_1$ is real (corresponding to the unique $G_2$-invariant three-form). 
We conclude that the warp factor is constant and $F=0$, so that the Killing spinor 
equation for $\chi_1$ (\ref{spinoreqns}) leads to weak $G_2$ holonomy and hence an 
Einstein metric. The second Killing spinor $\chi_2$ (for which the analysis is essentially the same) then of course leads to 
a Sasaki-Einstein manifold.

Alternatively, if $\|\xi\|=1$ then we immediately have $\Real S^*L=0$ by computing 
the square length of the latter using (\ref{Fierz}). But since 
also $a=-c|S|^2$ follows from linear dependence, we also have the additional 
relation $\Imag S^* L = |S|^2 K$ from (\ref{depend}). There is thus only 
one linearly independent vector, as one expects since we must have an $SU(3)$ structure.
Using the exterior derivatives of the one-form bilinears one can then show that where $S$ is non-zero 
we have that $K$ is closed, $\diff K=0$ (recall that $K$ is Killing in any case, so this implies that 
$K$ is parallel). 
By contracting $K$ into the bilinear equation for $\diff K$ and making use of a Fierz identity 
one then proves that $\diff\Delta=0$. Given that $\|\xi\|^2=|S|^2+\zeta^2 =1$ by 
assumption, this immediately implies that $S$ is constant, and hence that $L=0$. 
But then all vectors are identically zero, and we have a contradiction. Thus 
it must be that $S=0$ and we hence reduce to the previous case, 
which implies that $Y_7$ is Sasaki-Einstein with $F=0$ and $\Delta$ constant.

\section{The case $m=0$, $\mathrm{Im}\left[\bar\chi_1\chi_2\right]\neq 0$}
\label{app:m0}

In section \ref{pre} we noted that when $m=0$ we can no longer  conclude 
that equation (\ref{gammazero}) holds. In this appendix we study the case $m=0$ but 
$\mathrm{Im}\left[\bar\chi_1\chi_2\right]$ not being identically zero, in particular 
showing that there are no regular solutions in this class. 
Note this is different from the class of $m=0$ geometries discussed in 
section \ref{m0limit}, and cannot be obtained by taking the $m\rightarrow 0$ limit 
of the general $m\neq0$ equations in the main text. 

We begin by defining
\bea
h &\equiv & \mathrm{Im}\left[\bar\chi_1\chi_2\right]~,
\eea
which is a function on $Y_7$. Equation (\ref{K4eqn}) now becomes
\bea
\mathrm{Im}\, K &=& \frac{1}{2}\diff h~,
\eea
while the imaginary part of equation (\ref{Lie}) reads
\bea
\nabla_{(m} (\mathrm{Im}\, K)_{n)} &=& - 2h g_{7\, mn}~.
\eea
Combining the last two equations gives
\bea\label{obataeqn}
\nabla_m\nabla_n h &=& -t^2 h g_{7\, mn}~,
\eea
where $t=2$. Notice that $\mathrm{Im}\, K$ is a particular type of gradient conformal Killing vector.
Equation (\ref{obataeqn}) was studied by Obata in \cite{Obata}. In particular, 
he proved that if a complete Riemannian manifold of dimension $d\geq 2$ admits 
a non-constant function $h$ satisfying (\ref{obataeqn}), where $t$ is (without loss of generality) a positive constant, 
then it is necessarily isometric to a round sphere of radius $1/t$. Thus 
we immediately conclude that if $h$ is not identically zero, 
$Y_7$ is isometric to the round $S^7$ with radius $1/2$.

Now as in section \ref{m0limit}, the Bianchi identity and equation of motion 
for $F$ imply that $F$ is harmonic on the conformally rescaled manifold 
$(Y_7,\tilde{g}_7)$, where $\tilde{g}_7=\ex^{-6\Delta}g_7$. But in the case 
at hand, $Y_7=S^7$ and the Hodge theorem implies there are no harmonic 
four-forms since $H^4(S^7;\R)=0$. Thus for a non-singular solution in fact $F=0$,
 and hence the M-theory four-form $G=0$. The equation of motion (\ref{EOM})
then implies that the eleven-dimensional spacetime must be Ricci-flat, but this 
is a contradiction.

\end{document}